\documentclass[aps,prd,nofootinbib,12pt]{revtex4}
\usepackage{latexsym,amsmath,amssymb,amsbsy,mathbbol}
\usepackage{epsfig}

\newcommand{\bsym}{\boldsymbol}
\newcommand{\wtil}{\widetilde}
\newcommand{\ovl}{\overline}

\begin{document}

\title{Strange quarks in quenched twisted mass lattice QCD}

\author{Abdou M. Abdel-Rehim and Randy Lewis}

\affiliation{Department of Physics, University of Regina, Regina, SK,
             Canada, S4S 0A2}

\author{R. M. Woloshyn and Jackson M. S. Wu}

\affiliation{TRIUMF, 4004 Wesbrook Mall, Vancouver, BC, Canada, V6T 2A3}

\begin{abstract}
Two twisted doublets, one containing the up and down quarks and the other
containing the strange quark with an SU(2)-flavor
partner, are used for studies in the meson
sector.  The relevant chiral perturbation theory is presented, and
quenched QCD simulations (where the partner of the strange quark is not active) 
are performed.  Pseudoscalar meson
masses and decay constants are computed; the vector
and scalar mesons are also discussed.  A comparison is made to the case of
an untwisted strange quark, and some effects due to quenching,
discretization, and the definition of maximal twist are explored.
\end{abstract}

\maketitle

\section{Introduction}\label{sec:intro}

Twisted mass lattice QCD (tmLQCD) is a variation on the Wilson action
--- essentially a chiral rotation of quark flavor doublets, acting on quark
mass terms relative to Wilson terms in the action ---
which produces two desirable features: the removal of unphysical zero
modes in quark propagators\cite{Fetal01} and the elimination of $O(a)$
artifacts (where $a$ denotes lattice spacing) at maximal twist\cite{FR03}.
A number of numerical simulations have been performed for both quenched and
dynamical tmLQCD (for a recent review, see Ref.~\cite{Shindler}).
As well, the chiral perturbation theory for tmLQCD (tm$\chi$PT) 
has been developed. It differs from continuum $\chi$PT by 
discretization effects and is required for the extrapolation of 
tmLQCD data.
The effective theory has also played a vital role in
understanding various aspects of tmLQCD such as
$O(a)$ improvement, the phase diagram, and the relationships
between various definitions of maximal
twist\cite{MunsterSchmidt,Munster,Scorzato,SW04,AokiBar2004,SW05,Sharpe,
AokiBar2005}.

With an interest in the phenomenology of hadrons built of $u$, $d$ and $s$
quarks, our goal in this paper is to explore the usefulness of tmLQCD
and tm$\chi$PT as applied to strange hadrons. There is no unique way to
introduce the $s$ quark into the calculation. The method used here, to consider
a pair of quark doublets $(u,d)$ and $(``c",s)$, is similar to the proposal
of Pena {\it et al.}\cite{Pena}. For the quenched simulations considered here
the partner of the $s$ quark does not play an active role and should not be
thought of as the physical charm quark. In this work no mass splitting is 
introduced within either doublet. The construction of the 
corresponding tm$\chi$PT formalism is a straightforward 
generalization of the published one-doublet formalism\cite{SW04,SW05}.

As noted above, applying a relative chiral twist has some valuable 
consequences but there are also some less desirable features that 
have to be dealt with. The tmLQCD action violates parity conservation
so, in general, correlation functions contain contributions from states 
of both parities. Parity mixing can complicate, in particular, the extraction
of matrix elements but this can be ameliorated by appropriate tuning of the
twist angles. The tmLQCD action also breaks the flavor symmetry.
For the version of tmLQCD used in this work the members of the quark 
doublets are degenerate in mass but are distinguished by having 
opposite chiral twists. This can lead to mass splittings 
within hadron isospin multiplets. It will be seen that charged and
neutral kaons can acquire a mass-squared splitting which is roughly 
proportional to $a^2$.

To optimize the elimination of $O(a)$ lattice discretization errors
one has to tune the chiral twist angles\cite{FR03}. There is not a unique
way to achieve maximal twist as has been discussed from the point of
view of both effective theory\cite{SW04,SW05,Sharpe,AokiBar2004} 
and simulation\cite{Farchioni04,ALW,Jansen1}.
A standard method for defining maximal twist uses a tuning
procedure which involves the correlators of the first two isospin 
components of vector and axial operators with the pseudoscalar 
density\cite{Far05,Farchioni04,SW05}.  
Using two variations of this method, we examine the
mixing between the third isospin components of scalar and pseudoscalar
correlators. Ideally one would like to have a tuning to 
maximal twist which would banish the physical
pseudoscalar meson from appearing in the wrong parity correlator;
the scalar meson with its quenched $\eta^\prime\pi^0$ contribution would
similarly be banished from the other parity correlator. This is seen 
not to happen in our simulations.
The mixings observed in actual simulations represent higher order
discretization effects which differentiate between vector-axial tuning and
scalar-pseudoscalar tuning.

In this work, we mainly use maximal twist in the doublet containing the 
strange quark as well as in the $(u,d)$ doublet.
An alternative procedure would be to
set the twist angle for the strange quark to zero or equivalently for the
quenched theory to consider a twisted $(u,d)$ doublet and a flavor-singlet
Wilson strange quark. The latter approach may be a viable one for doing
full dynamical simulations. The twisted and untwisted strange quark
actions lead to different patterns of parity mixing and flavour symmetry
breaking at non-vanishing lattice spacing. We present some results obtained 
with an untwisted strange quark action to illustrate some of these differences.

At this point it is worth noting that there exist even other approaches 
for dealing with the strange quark. The proposal of Frezzotti and Rossi
in Ref.~\cite{FRnondegen} allows for a nondegenerate doublet in a way which
is suitable for dynamical simulations\cite{Farchioni05}.
In the limit where quark masses are degenerate within each doublet, it is
equivalent to the scheme used in this paper. However, for nondegenerate quarks,
twist and quark-mass splitting are associated with different flavor 
transformation generators. The fermion action contains terms which
mix flavors so that flavor symmetry breaking effects would be more 
complicated to deal with in simulations and in the effective theory than for 
the tmLQCD action considered in this work. A further example is
Ref.~\cite{dimop} where options for tmLQCD chosen to facilitate 
the calculation of the so-called kaon bag parameter are discussed.

In addition to meson masses, the pseudoscalar meson decay constants are
also considered.
With quark masses fixed by physical meson masses, the decay constants
$f_\pi$ and $f_K$ become absolute predictions, and are shown to compare
favorably with previous quenched simulations using other actions.
All results are consistent with tm$\chi$PT.

The remainder of the article is organized as follows.
Section~\ref{sec:ECL} defines the effective chiral Lagrangian with two
twisted flavor doublets, and
Sec.~\ref{sec:tmChPT} uses that Lagrangian to derive expressions involving the
pseudoscalar masses and decay constants.
Section \ref{sec:details} presents the tmLQCD action and explains the parameter
choices for our numerical simulations, then
Sec.~\ref{sec:pseudoscalar} discusses results obtained for
the pseudoscalar and vector mesons.
Scalar-pseudoscalar mixing is studied in Sec.~\ref{sec:scalar}, and
a direct comparison to kaons built from untwisted strange quarks is
given in Sec.~\ref{sec:Wilson}.  Section~\ref{sec:outlook} contains the
conclusions of our work.  Details of currents and densities in tm$\chi$PT
are collected in the Appendix.

\section{The effective chiral Lagrangian}\label{sec:ECL}

To build four-flavor tm$\chi$PT, we begin from tmLQCD
with two quark doublets,
\begin{equation}\label{defdoublets}
\psi_l=\begin{pmatrix} u \\ d \end{pmatrix} \,, \qquad
\psi_h=\begin{pmatrix} c \\ s \end{pmatrix} \,,
\end{equation}
referred to as the light and heavy doublets respectively. Note that the choice 
of flavor labels is a convention; in
Ref.~\cite{Pena} for example, a different choice is made.
In this work each doublet is taken to be degenerate,
so the $c$ quark, which is not active in any of our quenched tmLQCD 
simulations, should not be viewed as the physical charm quark.
Pena {\it et al.}\cite{Pena} discuss the extension of this case to the
case of a nondegenerate doublet where the quark-mass splitting is aligned with
the twist, preserving the favorable feature of no flavor mixing.
The fermion determinant 
does not remain real under this generalization so this would not 
lead a suitable action for nonquenched simulations.  However, this action may
still be useful for valence quarks in a mixed action scenario as discussed,
for example, in the context of tmLQCD in Ref.~\cite{mixedaction}.

In the so-called ``twisted basis''\cite{Fetal01,FR03}, the two-doublet lattice 
action is simply a block-diagonal version of two copies of the one-doublet theory
(the form of which can be found in Refs.~\cite{Fetal01,FR03}):
\begin{align} \label{E:action}
S^L_F &= a^4\sum_{x} \bar{\Psi}(x)
\Big[\frac{1}{2} \sum_{\mu} \gamma_\mu (\nabla^\star_\mu + \nabla_\mu)
- \frac{a}{2} \sum_{\mu} \nabla^\star_\mu \nabla_\mu 
+ \boldsymbol{m_0} + i \gamma_5 \boldsymbol{\mu_0} \Big] \Psi(x) \,,
\end{align}
where $\nabla_\mu$ and $\nabla^\star_\mu$ are the usual covariant forward and
backward lattice derivatives respectively, and
\begin{equation}
\Psi =
\begin{pmatrix}
\psi_l \\
\psi_h
\end{pmatrix} \,, \qquad
\bsym{m_0} = 
\begin{pmatrix}
m_{l,\,0}\mathbb{1}_2 & 0 \\
0 & m_{h,\,0}\mathbb{1}_2
\end{pmatrix} \,, \qquad
\bsym{\mu_0} = 
\begin{pmatrix}
\mu_{l,\,0}\tau_3 & 0 \\
0 & \mu_{h,\,0}\tau_3
\end{pmatrix} \,,
\end{equation}
with $\mathbb{1}_n$ the $n$-by-$n$ identity matrix. The matrix $\tau_3$ acts in 
(two-)flavor space and is normalized so that $\tau_3^2 = \mathbb{1}_2$. The 
parameters $m_{p,\,0}$ and $\mu_{p,\,0}$ are the normal 
bare and twisted masses respectively, with $p = l,\,h$.

Applying the now familiar two-step procedure of Ref.~\cite{SS98}, an effective 
chiral Lagrangian describing the low energy physics of tmLQCD with two degenerate 
quark doublets can be built as a straightforward generalization of the one-doublet 
case detailed in Refs.~\cite{RupakShoresh,BarRupakShoresh1,SW04,SW05}.  From
a similar analysis described in 
Ref.~\cite{SW04}, the form of the effective continuum Lagrangian at the quark level 
is found to be identical to that in the one-doublet case:
\begin{align} \label{E:contL}
\mathcal{L}_{\rm eff} &= \mathcal{L}_g + 
\bar{\Psi} (D \!\!\!\!/ + \bsym{m} + i\gamma_5\,\bsym{\mu}) \Psi 
+ b_{SW}\,a\,\bar{\Psi} i\sigma_{\mu\nu}F_{\mu\nu} \Psi + O(a^2) \,,
\end{align}
where $\mathcal{L}_g$ is the continuum gluon Lagrangian, and the physical normal and 
twisted mass parameters, $\bsym{m}$ and $\bsym{\mu}$, are defined analogously as in 
the one-doublet case:
\begin{eqnarray} 
\bsym{m} &\equiv& \begin{pmatrix} m_l & 0 \\ 0 & m_h \end{pmatrix}
= \begin{pmatrix} Z_{m,\,l}(m_{l,\,0}-\wtil{m}_{c,l}) & 0 \\
   0 & Z_{m,\,h}(m_{h,\,0}-\wtil{m}_{c,h}) \end{pmatrix} \,, \\
\bsym{\mu} &\equiv&
\begin{pmatrix}
\mu_l\tau_3 & 0 \\
0 & \mu_h\tau_3
\end{pmatrix}
= \begin{pmatrix}
Z_{\mu,\,l}\mu_{l,\,0}\tau_3 & 0 \\
0 & Z_{\mu,\,h}\mu_{h,\,0}\tau_3
\end{pmatrix}
= \begin{pmatrix}
Z_{P,\,l}^{-1}\mu_{l,\,0}\tau_3 & 0 \\
0 & Z_{P,\,h}^{-1}\mu_{h,\,0}\tau_3
\end{pmatrix} \,,
\end{eqnarray}
with $Z_{P,\,l}$ and $Z_{P,\,h}$ being the matching factors for the
pseudoscalar density.  The quantities $\wtil{m}_{c,\,l}$ and $\wtil{m}_{c,\,h}$
are the critical masses, aside from an $O(a)$ shift (see
Refs.~\cite{RupakShoresh,BarRupakShoresh1,SW04,SW05} and 
discussions below). Lattice symmetries forbid additive renormalization of 
$\mu_{l,\,0}$ and $\mu_{h,\,0}$.
As an aside, we note that symmetries also cause the ultraviolet divergent parts
of $\wtil{m}_{c,\,l}$ and $\wtil{m}_{c,\,h}$ to be identical.  One can choose
a definition of maximal twist (it will be called method (ii) in
Sec.~\ref{sec:details})
for which $\wtil{m}_{c,\,l}=\wtil{m}_{c,\,h}$, but here we do not restrict
the discussion to that special case.

Working to NLO in the power counting scheme,
\begin{equation}\label{powercounting}
m_l \sim m_h \sim \mu_l \sim \mu_h \sim p^2 \sim a\Lambda_{QCD}^2 \,, 
\end{equation}
the effective chiral Lagrangian found from matching reads
\begin{align} \label{E:chiralL}
\mathcal{L}_\chi &= 
 \frac{f^2}{4}\mathrm{Tr}(D_\mu\Sigma D_\mu\Sigma^\dagger)
-\frac{f^2}{4}\mathrm{Tr}(\chi^{\dagger}\Sigma + \Sigma^\dagger\chi) 
-\frac{f^2}{4}\mathrm{Tr}(\hat{A}^{\dagger}\Sigma + \Sigma^\dagger\hat{A}) 
\notag \\ 
&\quad
- L_1\big[\mathrm{Tr}(D_\mu\Sigma D_\mu\Sigma^\dagger)\big]^2
- L_2\mathrm{Tr}(D_\mu\Sigma D_\nu\Sigma^\dagger)
     \mathrm{Tr}(D_\mu\Sigma D_\nu\Sigma^\dagger) 
- L_3\mathrm{Tr}\big[(D_\mu\Sigma D_\mu\Sigma^\dagger)^2\big]
\notag \\
&\quad 
+ L_4\mathrm{Tr}(D_\mu\Sigma^\dagger D_\mu\Sigma)
     \mathrm{Tr}(\chi^{\dagger}\Sigma + \Sigma^\dagger\chi)
+ L_5\mathrm{Tr}\left[(D_\mu\Sigma D_\mu\Sigma^\dagger)
                      (\chi\Sigma^\dagger + \Sigma\chi^\dagger)\right]
\notag \\ 
&\quad
- L_6\big[\mathrm{Tr}(\chi^{\dagger}\Sigma + \Sigma^\dagger\chi)\big]^2
- L_7\big[\mathrm{Tr}(\chi^{\dagger}\Sigma - \Sigma^\dagger\chi)\big]^2
- L_8\mathrm{Tr}\big[(\chi^{\dagger}\Sigma + \Sigma^\dagger\chi)^2\big]
\notag \\ 
&\quad
+ i L_9\mathrm{Tr}(L_{\mu\nu} D_\mu\Sigma D_\nu\Sigma^\dagger +
                   R_{\mu\nu} D_\mu\Sigma^\dagger D_\nu\Sigma)
+ L_{10}\mathrm{Tr}(L_{\mu\nu}\Sigma R_{\mu\nu}\Sigma^\dagger) 
\notag \\
&\quad
+ W_4\mathrm{Tr}(D_\mu\Sigma^\dagger D_\mu\Sigma)
     \mathrm{Tr}(\hat{A}^{\dagger}\Sigma + \Sigma^{\dagger}\hat{A})
+ W_5\mathrm{Tr}\big[(D_\mu\Sigma D_\mu\Sigma^\dagger)
                     (\hat{A}\Sigma^\dagger + \Sigma\hat{A}^\dagger)\big]
\notag \\
&\quad
- W_6\mathrm{Tr}(\chi^{\dagger}\Sigma + \Sigma^\dagger\chi) 
     \mathrm{Tr}(\hat{A}^{\dagger}\Sigma + \Sigma^{\dagger}\hat{A})
- W_6'\big[\mathrm{Tr}(\hat{A}^{\dagger}\Sigma + \Sigma^{\dagger}\hat{A})\big]^2
\notag \\
&\quad
- W_7\mathrm{Tr}(\chi^{\dagger}\Sigma - \Sigma^\dagger\chi)  
     \mathrm{Tr}(\hat{A}^{\dagger}\Sigma - \Sigma^{\dagger}\hat{A})
- W_7'\big[\mathrm{Tr}(\hat{A}^{\dagger}\Sigma - \Sigma^{\dagger}\hat{A})\big]^2
\notag \\
&\quad
- W_8\mathrm{Tr}\big[(\chi^{\dagger}\Sigma +\Sigma^\dagger\chi)  
                     (\hat{A}^{\dagger}\Sigma + \Sigma^{\dagger}\hat{A})\big]
- W_8'\mathrm{Tr}\big[(\hat{A}^{\dagger}\Sigma + \Sigma^{\dagger}\hat{A})^2\big]
\notag \\
&\quad
+ W_{10}\mathrm{Tr}(D_\mu\hat{A}^\dagger D_\mu\Sigma + D_\mu\Sigma^\dagger D_\mu\hat{A}) 
\notag \\
&\quad 
+ H_1\mathrm{Tr}(L_{\mu\nu}L_{\mu\nu}+R_{\mu\nu}R_{\mu\nu})
- H_2\mathrm{Tr}(\chi^\dagger\chi)
- H_2'\mathrm{Tr}(\hat{A}^\dagger \chi+ \chi^\dagger \hat{A})
- H_3\mathrm{Tr}(\hat{A}^\dagger \hat{A}) \,,
\end{align}
where the $\Sigma$ field is now $SU(4)$ matrix-valued, and transforms under 
the chiral group $SU(4)_L \times SU(4)_R$.
Note that $\mathcal{L}_\chi$ has basically the same form as that in the $SU(2)$ 
theory\cite{RupakShoresh,BarRupakShoresh1,SW04,SW05},
except that terms linearly dependent under $SU(2)$ are no 
longer so under $SU(4)$.

The quantities $\chi$ and $\hat{A}$ are spurions for the quark masses and discretization errors, respectively\cite{BarRupakShoresh2}.
At the end of the analysis they are set to the constant values
\begin{equation}
\chi \longrightarrow 2B_0(\bsym{m} + i\bsym{\mu}) \,,
\qquad\qquad 
\hat{A} \longrightarrow 2W_0\,a\,\mathbb{1}_4 \,,
\end{equation}
where $B_0$ and $W_0$ are unspecified constants having dimensions
[mass] and [mass$^3$] respectively.
Notice that $\hat{A}$ involves a single flavour-independent Pauli term for
both doublets.

The discretization effect due to the Pauli term, {\it i.e.} the term containing
$b_{SW}$ in Eq.~\eqref{E:contL}, can be included non-perturbatively 
as in Ref.~\cite{SW05} by using the shifted spurion $\chi' \equiv \chi + \hat{A}$,
which corresponds at the quark level to a redefinition of the normal quark mass from 
$m$ to
\begin{equation}\label{E:m'def}
m'_p \equiv m_p + a W_0/B_0 \,\qquad p = l,\,h \,.
\end{equation}
This shift in turn corresponds to an $O(a)$ correction to the critical mass, so that 
it becomes 
\begin{equation}
m_{c,p} = Z_{m,\,p}\wtil{m}_{c,p} - a W_0/B_0 \,\qquad p = l,\,h \,.
\end{equation}
In terms of $\chi'$, the chiral Lagrangian, Eq.~\eqref{E:contL},
 can be written as:
\begin{align} \label{E:chiralL2}
\mathcal{L}_\chi &= 
 \frac{f^2}{4}\mathrm{Tr}(D_\mu\Sigma D_\mu\Sigma^\dagger)
-\frac{f^2}{4}\mathrm{Tr}(\chi'^{\dagger}\Sigma + \Sigma^\dagger\chi') 
\notag \\ 
&\quad
- L_1\big[\mathrm{Tr}(D_\mu\Sigma D_\mu\Sigma^\dagger)\big]^2
- L_2\mathrm{Tr}(D_\mu\Sigma D_\nu\Sigma^\dagger)
     \mathrm{Tr}(D_\mu\Sigma D_\nu\Sigma^\dagger) 
- L_3\mathrm{Tr}\big[(D_\mu\Sigma D_\mu\Sigma^\dagger)^2\big]
\notag \\
&\quad 
+ L_4\mathrm{Tr}(D_\mu\Sigma^\dagger D_\mu\Sigma)
     \mathrm{Tr}(\chi'^{\dagger}\Sigma + \Sigma^\dagger\chi')
+ L_5\mathrm{Tr}\left[(D_\mu\Sigma D_\mu\Sigma^\dagger)
                      (\chi'\Sigma^\dagger + \Sigma\chi'^{\dagger})\right]
\notag \\ 
&\quad
- L_6\big[\mathrm{Tr}(\chi'^{\dagger}\Sigma + \Sigma^\dagger\chi')\big]^2
- L_7\big[\mathrm{Tr}(\chi'^{\dagger}\Sigma - \Sigma^\dagger\chi')\big]^2
- L_8\mathrm{Tr}\big[(\chi'^{\dagger}\Sigma + \Sigma^\dagger\chi')^2\big]
\notag \\ 
&\quad
+ i L_9\mathrm{Tr}(L_{\mu\nu} D_\mu\Sigma D_\nu\Sigma^\dagger +
                   R_{\mu\nu} D_\mu\Sigma^\dagger D_\nu\Sigma)
%+ L_{10}\mathrm{Tr}(L_{\mu\nu}\Sigma R_{\mu\nu}\Sigma^\dagger) 
\notag \\
&\quad
+ \wtil{W}_4\mathrm{Tr}(D_\mu\Sigma^\dagger D_\mu\Sigma)
            \mathrm{Tr}(\hat{A}^{\dagger}\Sigma + \Sigma^{\dagger}\hat{A})
+ \wtil{W}_5\mathrm{Tr}\big[(D_\mu\Sigma D_\mu\Sigma^\dagger)
                            (\hat{A}\Sigma^\dagger + \Sigma\hat{A}^\dagger)\big]
\notag \\
&\quad
- \wtil{W}_6\mathrm{Tr}(\chi'^{\dagger}\Sigma + \Sigma^\dagger\chi') 
            \mathrm{Tr}(\hat{A}^{\dagger}\Sigma + \Sigma^{\dagger}\hat{A})
- \wtil{W}_6'\big[\mathrm{Tr}(\hat{A}^{\dagger}\Sigma + \Sigma^{\dagger}\hat{A})\big]^2
\notag \\
&\quad
- \wtil{W}_7\mathrm{Tr}(\chi'^{\dagger}\Sigma - \Sigma^\dagger\chi')  
            \mathrm{Tr}(\hat{A}^{\dagger}\Sigma - \Sigma^{\dagger}\hat{A})
- \wtil{W}_7'\big[\mathrm{Tr}(\hat{A}^{\dagger}\Sigma - \Sigma^{\dagger}\hat{A})\big]^2
\notag \\
&\quad
- \wtil{W}_8\mathrm{Tr}\big[(\chi'^{\dagger}\Sigma +\Sigma^\dagger\chi')  
                            (\hat{A}^{\dagger}\Sigma + \Sigma^{\dagger}\hat{A})\big]
- \wtil{W}_8'\mathrm{Tr}\big[(\hat{A}^{\dagger}\Sigma + \Sigma^{\dagger}\hat{A})^2\big]
\notag \\
&\quad
+ W_{10}\mathrm{Tr}(D_\mu\hat{A}^\dagger D_\mu\Sigma + D_\mu\Sigma^\dagger D_\mu\hat{A}) 
\notag \\
&\quad 
%+ H_1\mathrm{Tr}(L_{\mu\nu}L_{\mu\nu}+R_{\mu\nu}R_{\mu\nu})
- H_2\mathrm{Tr}(\chi'^\dagger\chi')
- \wtil{H}_2'\mathrm{Tr}(\hat{A}^\dagger\chi'+ \chi'^\dagger \hat{A})
%- \wtil{H}_3\mathrm{Tr}(\hat{A}^\dagger\hat{A}) 
\,,
\end{align}
where terms that lead only to contact terms in correlation functions
and are hence not needed below, viz. the $L_{10}$, $H_1$ and $H_3$ terms,
have been dropped.  We have also introduced useful combinations
\begin{align}
\wtil{W}_i &= W_i - L_i \,, \quad i = 4,\,5 \notag \\
\wtil{W}_j &= W_j - 2L_j \,, \quad \wtil{W}_j'= W_j'- W_j + 2L_j \,, \quad
j = 6,\,7,\,8 \notag \\
\wtil{H}_2' &= H_2' - H_2 \,.
\end{align}

As noted in Ref.~\cite{SW05}, the $W_{10}$ term is redundant. It can be 
transformed away by the change of variables 
\begin{equation}
\delta\Sigma = \frac{2 W_{10}}{f^2}(\Sigma\hat{A}^\dagger\Sigma -\hat{A}) \,.
\end{equation}
This transforms the $W_{10}$ term into the $\wtil{W}_5$, $\wtil{W}_8$, and $\wtil{H}_2'$ 
terms with their coefficients shifted to $\wtil{W}_5 + W_{10}$, $\wtil{W}_8 + W_{10}/2$, 
and $\wtil{H}_2' - W_{10}$, respectively. All physical quantities must depend then only 
on these combinations and not on $W_{10}$, $\wtil{W}_5$, $\wtil{W}_8$, and $\wtil{H}_2'$ 
separately. We have kept the $W_{10}$ term because it provides a useful diagnostic in
tm$\chi$PT calculations.

\section{Chiral perturbation theory for generic small masses}\label{sec:tmChPT}

In this section, we work out the consequences of the $SU(4)$ effective chiral 
Lagrangian, Eq.~\eqref{E:chiralL2}, which generalizes the results of the $SU(2)$ theory
of Ref.~\cite{SW05}, and our focus will be on the masses and decay constants of
kaons: pseudoscalar mesons that involve both flavor doublets.
We work in the ``generic small mass'' regime defined by
\begin{equation}\label{GSM}
\Lambda_\chi^2 \gg M_h' \gtrsim M_l' \gtrsim 2W_0\,a \,,
\end{equation}
where $\Lambda_\chi = 4\pi f$ and
\begin{equation}
M_p' = 2B_0\sqrt{m_p^{'2} + \mu_p^2} \,, \qquad p = l,\,h \,.
\end{equation}
Note that $M_p'$ has dimension [mass$^2$].
In the analysis of our numerical
data, quenching effects will also be considered
(see Eqs.~(\ref{msqquenched1})--(\ref{msqquenched2})).

\subsection{The vacuum}\label{subsec:vacmass}

At LO the discretized Lagrangian retains its continuum form, so
the LO vacuum expectation value 
(VEV) of $\Sigma$ is that which cancels out the twists in the shifted mass matrix:
\begin{equation}
\langle 0|\Sigma|0 \rangle_{LO} \equiv \Sigma_0 \equiv 
\begin{pmatrix}
\exp(i\omega_{l,\,0}\tau_3) & 0 \\
0 & \exp(i\omega_{h,\,0}\tau_3)
\end{pmatrix} \,,
\end{equation}
where $\omega_{p,\,0}$ are defined by
\begin{equation}
c_{p,\,0} \equiv \cos(\omega_{p,\,0}) = 2B_0m'/M_p' \,, \qquad
s_{p,\,0} \equiv \sin(\omega_{p,\,0}) = 2B_0\mu_p/M_p' \,, \qquad
p = l,\,h \,.
\end{equation}
This provides one definition for the twist angles.

At NLO, the VEV of $\Sigma$ is realigned by a small amount from $\Sigma_0$. Defining
\begin{equation}
\langle 0|\Sigma|0 \rangle_{NLO} \equiv \Sigma_m \equiv 
\begin{pmatrix}
\exp(i\omega_{l,\,m}\tau_3) & 0 \\
0 & \exp(i\omega_{h,\,m}\tau_3)
\end{pmatrix} \,, \quad
\omega_{p,\,m} = \omega_{p,\,0} + \epsilon_p \,, \quad p = l,\,h \,,
\end{equation}
the shifts from LO, $\epsilon_{l,\,h}$, are found from minimizing the potential to be
\begin{equation}
\epsilon_p = -\frac{16W_0as_{p,\,0}}{f^2}\left\{
2\wtil{W}_6\,(M_l' + M_h')/M_p' + \wtil{W}_8 
+ \frac{4W_0a}{M_p'}
\big[\wtil{W}_6'\,(c_{l,\,0} + c_{h,\,0}) + \wtil{W}_8'\,c_{p,\,0}\big]
\right\} \,.
\end{equation}

Expanding about the VEV as in Ref.~\cite{SW05}, the physical pion fields are defined by
\begin{equation} \label{E:Sigmaphys}
\Sigma = \xi_m \Sigma_{ph} \xi_m \,, \quad
\xi_m = 
\begin{pmatrix}
\exp(i\omega_{l,\,m}\tau_3/2) & 0 \\
0 & \exp(i\omega_{h,\,m}\tau_3/2)
\end{pmatrix} \,, \quad
\Sigma_{ph} = \exp(i\Phi/f) \,,
\end{equation}
where $\Phi$ has the representation
\begin{align}\label{defPhi}
\frac{1}{\sqrt{2}}\,\Phi &\equiv 
\frac{1}{\sqrt{2}}\sum_{i = 1}^{15}\varphi_i\,\Lambda_i \notag \\
&=
\begin{pmatrix}
\frac{1}{\sqrt{2}}\,\pi^0 +\frac{1}{\sqrt{6}}\,\eta_8 +\frac{1}{\sqrt{12}}\,\eta_{15} & 
\pi^+  & \ovl{D}^0 & K^+       \\
\pi^-  & 
-\frac{1}{\sqrt{2}}\,\pi^0 +\frac{1}{\sqrt{6}}\,\eta_8 +\frac{1}{\sqrt{12}}\,\eta_{15} & 
 D^-   & K^0   \\
 D^0   & D^+ & -\frac{3}{\sqrt{12}}\,\eta_{15} & D^+_s \\
 K^-   & \ovl{K}^0 & D^-_s & -\frac{2}{\sqrt{6}}\,\eta_8 + \frac{1}{\sqrt{12}}\,\eta_{15}
\end{pmatrix} \,.
\end{align}
Our choice of the fifteen generators of $SU(4)$, $\Lambda_1,\ldots,\Lambda_{15}$,
differ slightly from the conventional ones. Here, all off-diagonal generators and the diagonal 
$\Lambda_3$ are the same as the conventional ones, but we choose the rest of the diagonal generators
to be such that the $(3,3)$ and $(4,4)$ entries of the diagonal $\Lambda_8$ and $\Lambda_{15}$ are
interchanged with respect to the conventional ones\footnote{Note that once the choice is made for 
$\Lambda_8$, $\Lambda_{15}$ is fixed by the normalization condition 
$\mathrm{Tr}(\Lambda_i\Lambda_j)=2\delta_{ij}$.}.
This maintains consistency with the ordering of quark fields ($u$, $d$, $c$, $s$), used throughout 
this work\footnote{Recall that $u$ and $c$ have a positive twist whereas $d$ and $s$ have a negative 
twist.}, and allows for the standard meson naming convention. In particular, we have for 
the flavor-diagonal components of $\Phi$:
\begin{eqnarray}
|\pi^0\rangle &=& \frac{1}{\sqrt{2}}\bigg(|u\bar{u}\rangle - |d\bar{d}\rangle
                  \bigg) \,, \\
|\eta_8\rangle &=& \frac{1}{\sqrt{6}}
\bigg(|u\bar{u}\rangle + |d\bar{d}\rangle - 2|s\bar{s}\rangle\bigg) \,, \\
|\eta_{15}\rangle &=& \frac{1}{\sqrt{12}}
\bigg(|u\bar{u}\rangle + |d\bar{d}\rangle + |s\bar{s}\rangle
 - 3|c\bar{c}\rangle\bigg) \,.
\end{eqnarray}
Note that the $c$ quark is mass-degenerate with the $s$ quark, so pairs
of $D$'s and $K$'s related by the interchange of $c$ and $s$ quarks are
mass-degenerate at LO in the chiral expansion.  By inserting the above expansion of 
$\Sigma$ into the chiral Lagrangian, Eq.~\eqref{E:chiralL2}, the Feynman rules for the $SU(4)$ 
theory can be straightforwardly obtained.

\subsection{Defining the twist angle}\label{subsec:twangle}

In the continuum, the twist angle for each degenerate doublet can be defined 
unambiguously by
\begin{equation}
\omega_p = \tan^{-1}(\mu_p/m_p) \,, \qquad p = l,\,h \,.
\end{equation}
Given the twist angles, the off-diagonal $SU(4)$ components of physical currents
and densities are related to their counterparts 
in the twisted basis by (with a ``hat'' denoting the physical basis)
\begin{align} \label{E:JCPhys2TM}
\hat{V}^a_\mu &= \cos(\Delta^a\omega)V^a_\mu + \eta_{ab}\sin(\Delta^a\omega)A^b_\mu
\,,
\notag \\
\hat{A}^a_\mu &= \cos(\Delta^a\omega)A^a_\mu + \eta_{ab}\sin(\Delta^a\omega)V^b_\mu 
\,,
\notag \\
\hat{S}^a &= \cos(\Sigma^a\omega)S^a + i\sin(\Sigma^a\omega)P^a \,,
\notag \\
\hat{P}^a &= \cos(\Sigma^a\omega)P^a + i\sin(\Sigma^a\omega)S^a \,,
\notag \\
a,\,b&\in\mathfrak{K}\backslash\mathfrak{D} \,, \qquad
\mathfrak{K} = \{1,\ldots,15\} \,, \qquad \mathfrak{D} = \{3,\,8,\,15\} \,,
\end{align}
where
\begin{equation}\label{defapp1}
\eta_{ab} = \left\{\begin{array}{rl}\pm 1, & {\rm for~} b = a \pm 1 \\
                                        0, & {\rm otherwise} \end{array}
      \right.
\end{equation}
and
\begin{equation}
\Delta^a\omega = \frac{1}{2}(\omega_{i_a} - \omega_{j_a}) \,, \qquad
\Sigma^a\omega = \frac{1}{2}(\omega_{i_a} + \omega_{j_a}) \,, \qquad i_a < j_a \,.
\end{equation}
The indices $i_a$ and $j_a$ are the row numbers of the non-zero entries of the $SU(4)$ 
generator $\Lambda_a$ that defines the particular flavor current or density, and
\begin{equation}\label{defapp3}
\omega_{1} = -\omega_{2} = \omega_{l} \,, \qquad
\omega_{3} = -\omega_{4} = \omega_{h} \,.
\end{equation}
For a thorough discussion of currents and densities, see
App.~\ref{app:currents}.  We note now that Eq.~\eqref{E:JCPhys2TM}
has the form of
the inverse transformation of the LO operator, Eq.~\eqref{E:LOJD}, except that here 
the twist angles are $\omega_p$ not $\omega_{p,\,m}$.

On a lattice, discretization errors mean that different definitions for
the twist angles will lead to 
observables that differ by $O(a)$.  In this work, we will define $\omega_p$ 
non-perturbatively as in Refs.~\cite{Far05,Farchioni04,SW04,ALW}
by enforcing the absence of
parity breaking in the physical basis.  In particular, we will enforce
\begin{equation} \label{E:NPwcond}
\langle\hat{V}^b_\mu(x)\hat{P}^a(y)\rangle = 0 \,, \qquad 
a,\,b\in\mathfrak{K}\backslash\mathfrak{D} \,.
\end{equation}
{}From the definitions in Eq.~\eqref{E:JCPhys2TM}, this condition gives
\begin{equation} \label{E:NPwdef}
\tan\omega_l \equiv 
\frac{\langle V^2_\mu(x)P^1(y)\rangle}{\langle A^1_\mu(x)P^1(y)\rangle} \,, \qquad
\tan\omega_h \equiv 
\frac{\langle V^{14}_\mu(x)P^{13}(y)\rangle}{\langle A^{13}_\mu(x)P^{13}(y)\rangle} \,.
\end{equation}
The results for $\omega_p$ depend on the distance $|x - y|$ at $O(a)$. We will enforce 
the condition in Eq.~\eqref{E:NPwcond} at long distance where the
single-meson contribution dominates. 
 
Evaluating Eq.~\eqref{E:NPwdef} using the results in App.~\ref{app:currents}, we find at LO 
$\omega_p = \omega_{p,\,m} = \omega_{p,\,0}$, since only $\hat{A}^a_{\mu,\,LO}$ and 
$P^a = \hat{P}^a$, $a\in\{1,13\}$, couple to the single-meson state. At NLO, the sole
non-trivial contributions surviving in the ratios of Eq.~\eqref{E:NPwdef} come from 
the $W_{10}$ term, just as in the one-doublet theory, and as in Ref.~\cite{SW05} we find
\begin{equation} \label{E:tanwp}
\tan\omega_p = \frac{\sin\omega_{p,\,m}}{\cos\omega_{p,\,m}+\delta} \,, \qquad
\delta = \frac{8W_0 a}{f^2} W_{10} \,, \qquad p = l,\,h \,.
\end{equation}

\subsection{Kaon masses and decay constants}\label{subsec:Kmass}
With the Feynman rules in hand, and the twist angles defined, we have now all that 
is needed to calculate pseudoscalar meson masses and matrix elements. At NLO, we find that the 
mass of the neutral kaon is given by
\begin{eqnarray}\label{E:mKsq}
m^2_{K^0} &=& \hat{M}'
+ \frac{\hat{M}'}{6}\Big\{
4\mathcal{I}[m_{\eta_8}^2] - \mathcal{I}[m_{\eta_{15}}^2]\Big\}
\notag \\
&&+ \frac{8}{f^2}\Big\{\hat{M}'^2\,(8L_6 - 4L_4 + 2L_8 - L_5) 
\notag \\
&&+ \hat{M}'a\,W_0\,(c_l + c_h)(8\wtil{W}_6 - 4\wtil{W}_4 + 2\wtil{W}_8 - \wtil{W}_5) 
\notag \\
&&+ 2W_0^2\,a^2\,\big[(4\wtil{W}_6' + \wtil{W}_8')(c_l + c_h)^2 - \wtil{W}_8'\,(s_l + s_h)^2
\big] \Big\} \,,
\end{eqnarray}
where
\begin{eqnarray}
\hat M' &=& (M_l'+M_h')/2 \,, \\
m_{\eta_8}^2 &=& (M_l' + 2M_h')/3 \,, \\
m_{\eta_{15}}^2 &=& (M_l' + 5M_h')/6 \,,
\end{eqnarray}
are respectively the LO $K$, $\eta_8$ and $\eta_{15}$ squared masses,
and $c_p = \cos\omega_p$, $s_p = \sin\omega_p$, which we can use instead of $\cos\omega_{p,\,m}$ and 
$\sin\omega_{p,\,m}$ respectively at the order we work.
Note that the usual continuum one-loop contribution\cite{GL85} 
appears in Eq.~\eqref{E:mKsq},
\begin{equation}
\mathcal{I}[m^2] 
= \frac{m^2}{32\pi^2 f^2}\ln\frac{m^2}{\Lambda_R^2} \,,
\end{equation}
with $\Lambda_R$ being the renormalization scale.
Note also that, as required, the neutral kaon mass depends only on the 
combination $2\wtil{W}_8 - \wtil{W}_5$ rather than on $\wtil{W}_8$ and
$\wtil{W}_5$ separately, and
that the kaon mass is automatically 
$O(a)$ improved at maximal twist ($c_p = 0$).

Flavor breaking in the kaon masses at NLO is given solely by the analytic
contribution, just as in the $SU(2)$ theory\cite{SW05}, and it reads
\begin{equation}\label{Kmassdiff}
m^2_{K^0} - m^2_{K^\pm} 
= -\frac{64}{f^2}W_0^2\,a^2 s_l s_h \wtil{W}_8' 
= -\wtil{W}_8'\,\frac{64}{f^2}\frac{W_0^2\,a^2\,\mu_l\,\mu_h}{M_l'\,M_h'} \,,
\end{equation}
where the second equality is derived using the fact that we can replace $s_p$ by $s_{p,0}$ 
at the order we are working.  

Before calculating Eq.~\eqref{Kmassdiff} explicitly, one might have
anticipated an expression that was quadruply suppressed in our power
counting scheme, Eq.~\eqref{powercounting}, due to the requirement of being
$O(a^2)$ and $O(\mu_l\mu_h)$.
However, Eq.~\eqref{Kmassdiff} shows that the $\mu_l\mu_h$ dependence
enters as a ratio with $M_l'M_h'$ and the squared kaon mass difference
is therefore nonzero already at NLO.

With the physical axial current defined in App.~\ref{app:currents}, the
$K^\pm$ decay constant to NLO is determined to be
\begin{align}\label{E:fK}
f_K = f\bigg\{1 &+ \frac{4}{f^2}\big[(4L_4 + L_5)\hat{M}'
+ W_0\,a\,(c_l + c_h)(4\wtil{W}_4 + \wtil{W}_5 + W_{10})\big] \notag \\
& -\frac{3}{4}\mathcal{I}[m_\pi^2] - \frac{3}{4}\mathcal{I}[m_{\eta_8}^2] 
- 2\mathcal{I}[m_K^2] - \frac{1}{2}\mathcal{I}[m_{D_s}^2]\bigg\} \,,
\end{align}
where $m_\pi^2 = M_l'$, $m_K^2 = m_D^2 = \hat{M}'$ and $m_{D_s}^2 = M_h'$
are the LO expressions for the
pion, kaon and $D_s$ squared masses respectively.
Our tm$\chi$PT conventions are such that $f_\pi\approx93$ MeV.
Note that the one-loop contributions 
from the pions and the $\eta_8$ are the same as in the continuum $SU(3)$ theory\cite{GL85}.
Flavor breaking effects enter first at $O(a^2)$, which is NNLO for decay constants. The 
above result shows that the decay constant depends only on the combination 
$\wtil{W}_5 + W_{10}$, and is automatically $O(a)$ improved at maximal twist.

\section{Simulation details}\label{sec:details}

We have performed quenched simulations using the action of Eq.~\eqref{E:action}.
The ensembles computed in Ref.~\cite{ALW}, containing 300 gauge configurations each at
$\beta=5.85$ and $\beta=6.0$, have subsequently been extended to include 600
configurations each\cite{ALWdublin}.  An additional ensemble at $\beta=6.2$ has also
been generated, again using a pseudo-heatbath algorithm which acts on all $SU(2)$
subgroups.
Quark propagators are obtained from a 1-norm quasi-minimal residual algorithm,
with periodic boundary conditions in all directions.
Simulation parameters are collected in Table~\ref{tab:parameters}.

To remove $O(a)$ errors, simulations must be done at maximal twist, and to
that end we employ Eq.~\eqref{E:NPwdef} implemented, following
Refs.~\cite{Far05,Farchioni04,ALW} by
\begin{eqnarray}\label{defangle}
\tan\omega_l
 &=& \frac{i\sum_{\vec x}\left<V_4^{1-i2}(\vec x,t)P^{1+i2}(0)\right>}
           {\sum_{\vec x}\left<A_4^{1-i2}(\vec x,t)P^{1+i2}(0)\right>} \,, \\
\tan\omega_h
 &=& \frac{i\sum_{\vec x}\left<V_4^{13-i14}(\vec x,t)P^{13+i14}(0)\right>}
           {\sum_{\vec x}\left<A_4^{13-i14}(\vec x,t)P^{13+i14}(0)\right>} \,.
\end{eqnarray}
where ${\cal O}^{a+ib}\equiv({\cal O}^a+i{\cal O}^b)/\sqrt{2}$.
This method of tuning separately at each twisted quark mass is only one
possible choice.  A nice summary of the situation in the
generic small mass regime (Eq.~\eqref{GSM}) that we are using
has been given by Sharpe in Ref.~\cite{Sharpe}.
In the language of that article, we are using method (i).  A variant,
called method (ii), extrapolates the results of method (i) to the point of
vanishing twisted mass, and then uses that definition of maximal twist for
all values of $\mu_{p,\,0}$ ($p=l,h$).  Another option, called method (iii),
relies on
twistings in the scalar-pseudoscalar sector rather than the
vector-axial sector.  Method (iv) assumes that
maximal twist can be sufficiently well defined by simply holding the
hopping parameter fixed at its critical value from the
untwisted Wilson theory.

As sketched in Fig.~1 of Ref.~\cite{Sharpe}, methods (i), (ii) and (iii)
are all acceptable non-perturbative definitions of maximal twist, and
are superior to method (iv).
Though most of our simulations use method (i), we will make frequent
comparisons with results from the $\chi$LF
collaboration\cite{flavorbreaking,scaling} using method (ii).
We will also present our own results from method (ii) when discussing
aspects of scalar correlators in Sec.~\ref{sec:scalar}.

Throughout this work, only local operators are used,
and error bars reported in graphs and tables are statistical
only.  All statistical uncertainties are obtained from the
bootstrap method with replacement, where the number of bootstrap ensembles is
three times the number of data points in the original ensemble.
Masses and decay constants are obtained from unconstrained three-state fits to
correlators, using all time steps except the source.
In the following sections, each discussion includes references to the relevant figures,
but we note here that the corresponding numerical values are collected in
Table \ref{tab:allnumbers}.

\section{Pseudoscalar and vector mesons}\label{sec:pseudoscalar}

\begin{figure}[tb]
\scalebox{0.38}{\includegraphics*[0cm,11mm][255mm,19cm]{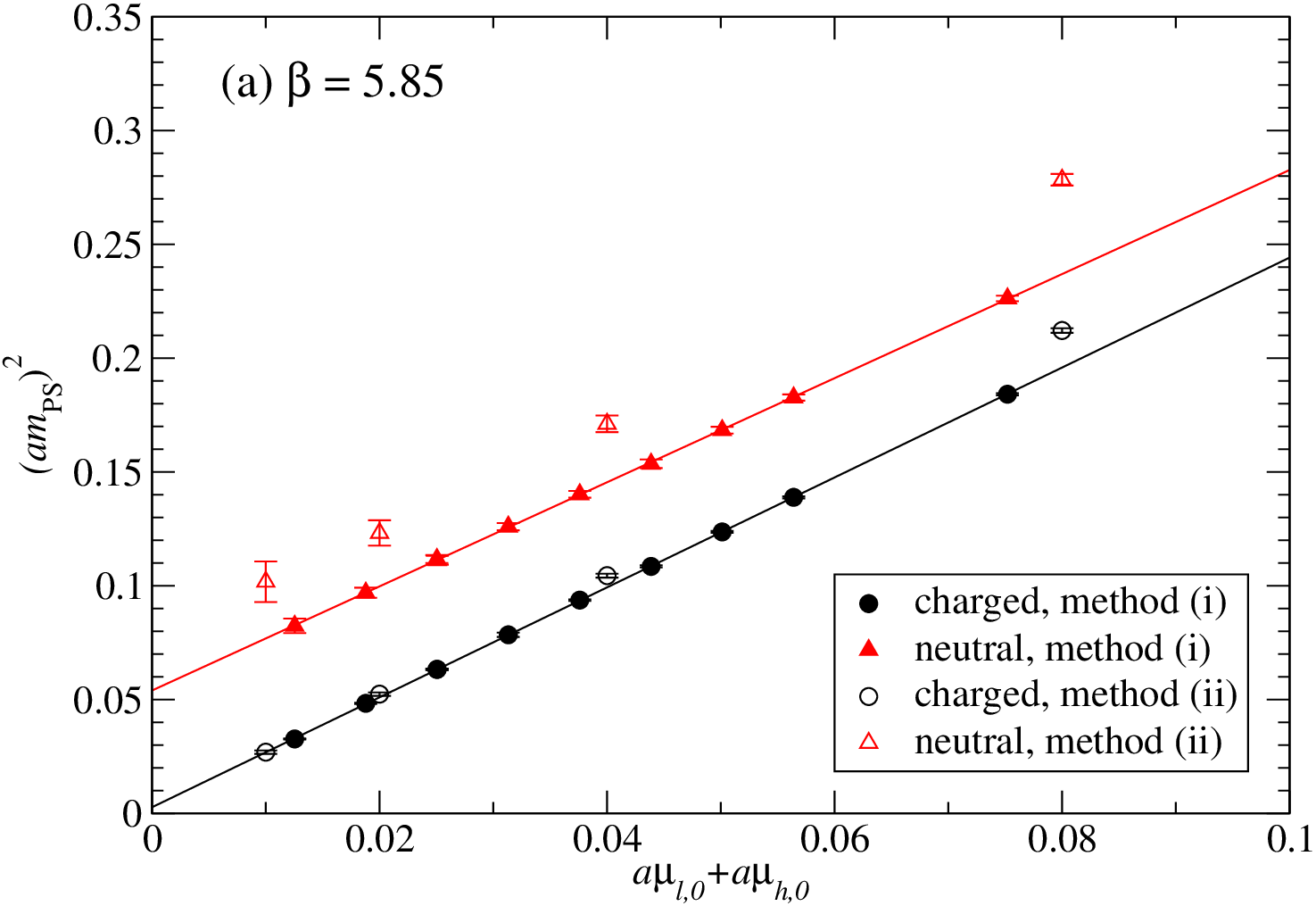}}
\scalebox{0.38}{\includegraphics*[0cm,11mm][255mm,19cm]{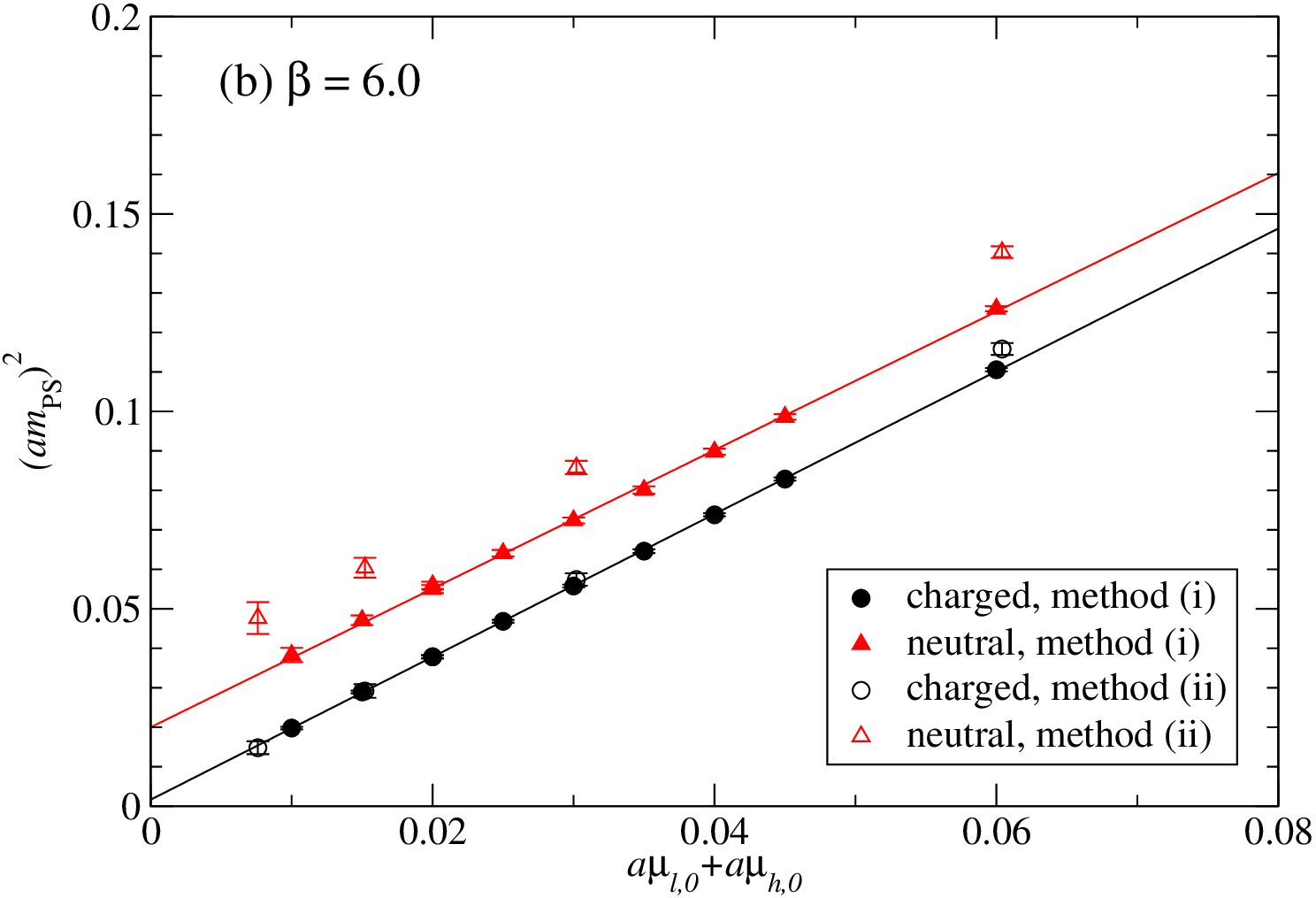}}
\scalebox{0.38}{\includegraphics*[0cm,11mm][255mm,19cm]{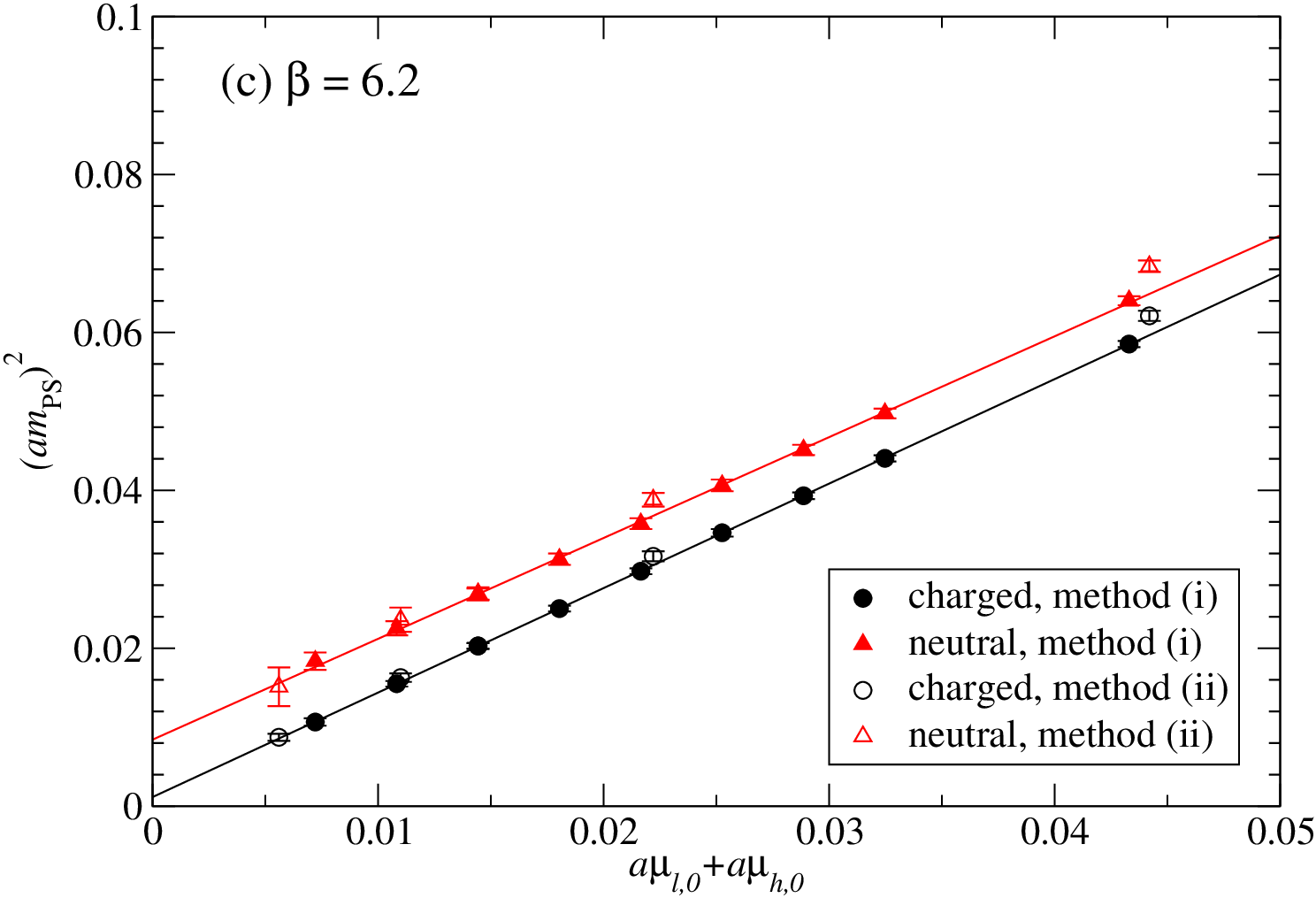}}
\caption{Pseudoscalar meson mass squared as a function of the sum of quark and
         antiquark twisted mass parameters.  Subscripts $l$ and $h$ indicate
         the light and heavy doublets.  Results labelled by method (i)
         are from the present work; results labelled by method (ii) are from
         Ref.~\protect\cite{flavorbreaking} and have equal masses for the
         quark and anti-quark.  Straight lines are linear fits
         to the data from method (i).}
\label{fig:mPSvsmumu}
\end{figure}

Masses of the charged and neutral kaons,
{\it i.e.} the ground state pseudoscalar mesons containing one $s$ (anti)quark from
the heavy doublet and one $u$ or $d$ (anti)quark from the light doublet, are
plotted in Fig.~\ref{fig:mPSvsmumu}.  Both twisted masses, $a\mu_{l,\,0}$ and
$a\mu_{h,\,0}$, take on all values from Table~\ref{tab:parameters} and the
normal mass term is tuned accordingly (also shown in
Table~\ref{tab:parameters}) to achieve maximal twist at each
particular twisted mass value.

These data are expected to be consistent with a quenched
version\cite{quench1,quench2,CPPACS} of
Eq.~\eqref{E:mKsq}, which at maximal twist reads
\begin{align}
m^2_{K^0} 
&= \hat{M}'\left[1-\delta_{\rm quench}\left(\ln\frac{M_l'}{(4\pi f)^2} 
  +\frac{M_h'}{M_h'-M_l'}\ln\frac{M_h'}{M_l'}\right)\right] \notag \\
&\quad +\frac{8\hat M'^2}{f^2}(8L_6 - 4L_4 + 2L_8 - L_5) 
        -\frac{64}{f^2}a^2W_0^2\wtil W_8' \label{msqquenched1} \\
&= \hat{M}' -\frac{64}{f^2}a^2W_0^2\wtil W_8' + O(M_p'^2)
 + O(\hat M'\delta_{\rm quench}) \,,
\qquad p=l,h \,, \\
m^2_{K^\pm} 
&= m^2_{K^0} + \frac{64}{f^2}a^2W_0^2\wtil W_8' \notag \\
&= \hat{M}' + O(M_p'^2) + O(\hat M'\delta_{\rm quench}) \,, \qquad p=l,h \,,
   \label{msqquenched2}
\end{align}
where $\delta_{\rm quench}$ is the standard coefficient for the
quenched logarithm.
As Fig.~\ref{fig:mPSvsmumu} indicates, linear fits in $\hat{M}'$ to the data
at each $\beta$ value yield excellent results, so $O(M_p'^2)$ and
$O(\hat M'\delta_{\rm quench})$ effects are largely unnecessary.  Notice in particular
that our linear fit for the charged kaon did lead to a nonzero (but small)
residual kaon mass at $\hat M'=0$,
thus reminding us that corrections to the linear form are important for such
details.  We have verified, for example, that the function
$A\hat M'+B\hat M'\ln\hat M'$ yields an equally excellent fit to our data,
and of course it enforces the absence of any residual mass at $\hat M'=0$.

Fig.~\ref{fig:mPSvsmumu} also reveals differences between our results
with method (i) and results from the $\chi$LF
collaboration\cite{flavorbreaking} using equal quark and antiquark masses
with method (ii).
For charged mesons, the method (i) mass difference is smaller than the
method (ii) difference, and as expected from tm$\chi$PT the chiral limits
appear to be very similar.
For neutral mesons, the chiral limits appear to be somewhat different on the
coarser
lattices, particularly at $\beta=6.0$, but become consistent at $\beta=6.2$.
We recall from Fig.~2 of Ref.~\cite{flavorbreaking} that the
$\chi$LF data at $\beta=6.0$ happen to be statistically above their fitted
$a^2$ extrapolation, so we see no essential disagreement among
any of the data sets displayed in our Fig.~\ref{fig:mPSvsmumu}.

\begin{figure}[tb]
\scalebox{0.45}{\includegraphics*[0cm,11mm][255mm,19cm]{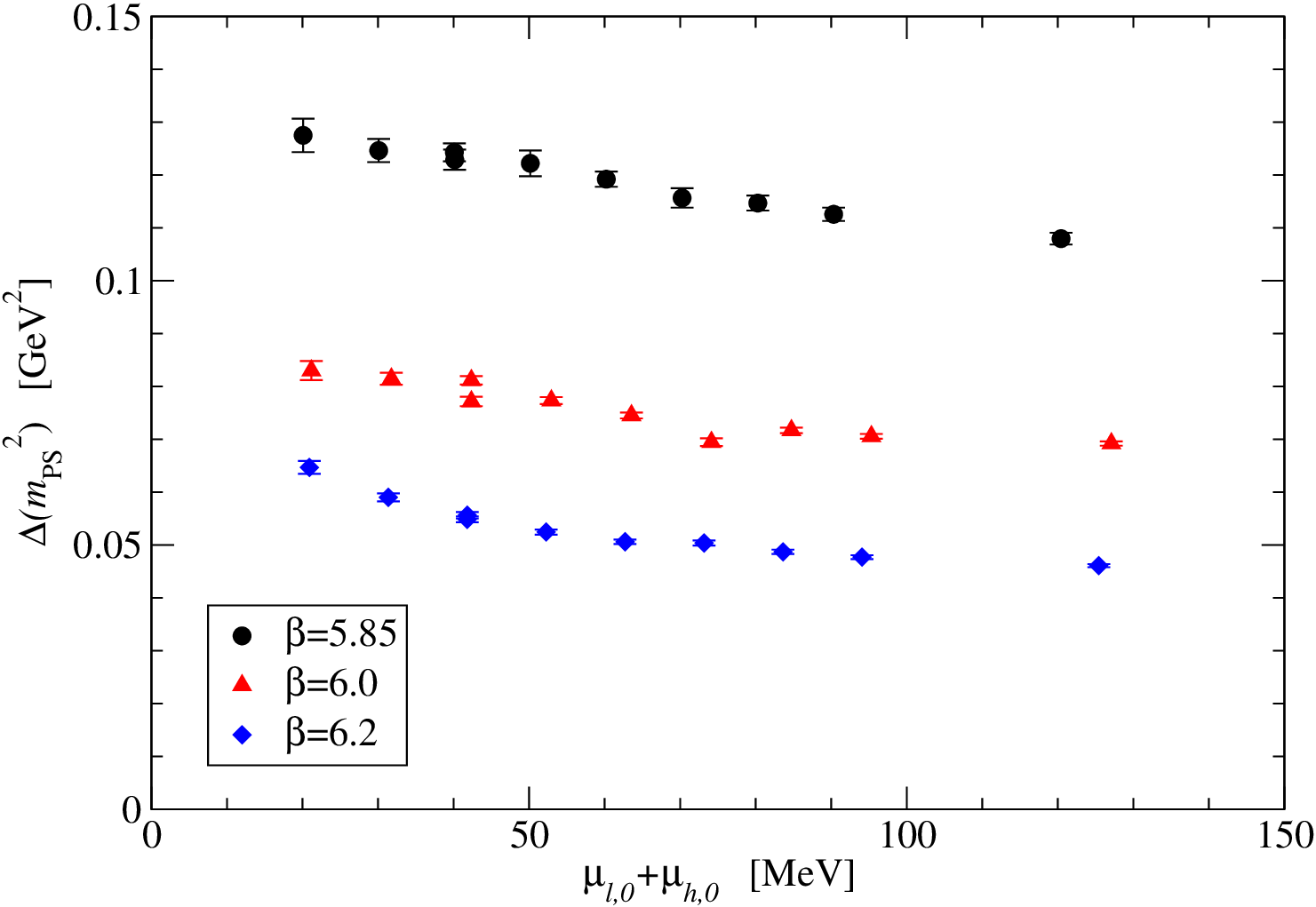}}
\caption{The difference between charged and neutral squared pseudoscalar meson
         masses as a function of the sum of quark and
         antiquark twisted masses.  Subscripts $l$ and $h$ indicate
         the light and heavy doublets.}
\label{fig:mPSdiffvsmqmq}
\end{figure}

The difference between the squared masses of charged and neutral kaons
is plotted directly in Fig.~\ref{fig:mPSdiffvsmqmq}, and is found to be
only mildly dependent on (twisted) quark mass over the range we are studying.
This implies that corrections to Eq.~\eqref{Kmassdiff}, arising from higher
orders in the tm$\chi$PT expansion, are small but noticeable.

\begin{figure}[tb]
\scalebox{0.45}{\includegraphics*[0cm,11mm][255mm,19cm]{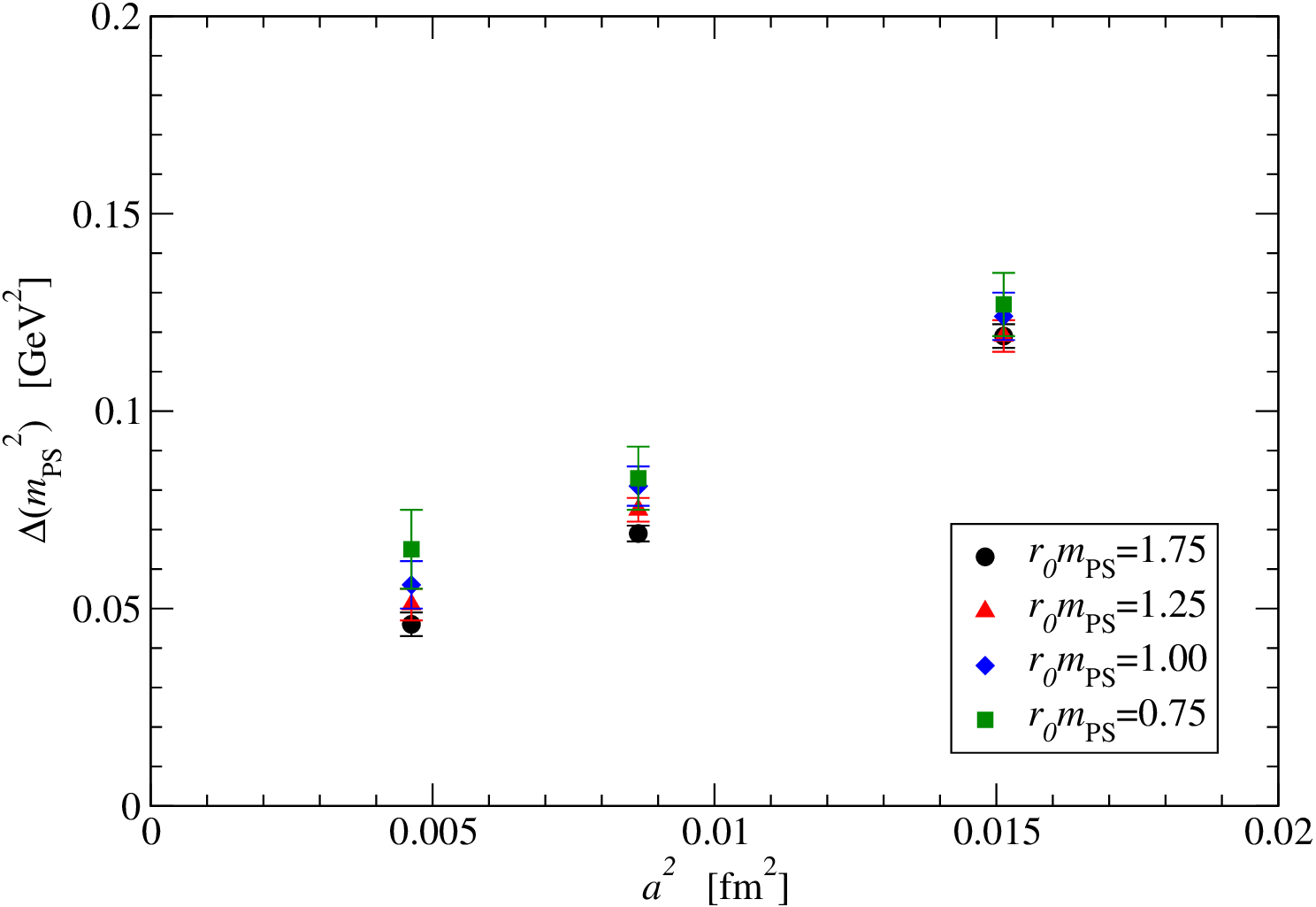}}
\caption{The difference between charged and neutral squared pseudoscalar meson
         masses as a function of squared lattice spacing, for selected
         values of the charged meson mass.}
\label{fig:mKsqvsasq}
\end{figure}

Fig.~\ref{fig:mKsqvsasq} shows the lattice spacing dependence of the squared
mass differences for four mass values that span the range of our available
data.  At leading-order in tm$\chi$PT, Eq.~\eqref{Kmassdiff} indicates that
this quantity should be independent of mass, linear in $a^2$ and vanishing
in the continuum limit.  Modulo the unknown higher order effects,
Fig.~\ref{fig:mKsqvsasq} is in reasonable agreement with these expectations.
In particular, the approximate mass independence is evident and the
dependence on $a^2$ is approximately linear, though a linear fit misses the
massless prediction at $a=0$ by a few (statistical) standard deviations.

It should be noted from Fig.~\ref{fig:mKsqvsasq} that even at our smallest
lattice spacing, the mass splitting of $m_{K^0}-m_{K^\pm}\sim50$ MeV is
significant relative to the kaon mass itself. However, in
terms of the difference of mass squared, our results are consistent 
with the pseudoscalar meson mass splittings in Ref.~\cite{flavorbreaking} and
compatible with the suggestion of Shindler\cite{Shindler} that flavor
breaking effects in tmLQCD are of a magnitude  
comparable to ``taste'' symmetry violations in pseudoscalar meson masses
observed with improved staggered fermions\cite{Aubin}.
The appearance of sizable lattice spacing effects like this have led some
authors to use a power counting scheme in which $O(a^2)$ effects are taken
to be LO rather than NLO\cite{Aoki,AokiBar2005}, but we will follow
Eq.~(\ref{powercounting}) throughout the present work.

The decay constant of a charged pseudoscalar meson can be obtained easily 
from the so-called indirect method\cite{FrezSint,Bietenholz},
\begin{equation}\label{indirect}
f_{\rm PS} = \left.\left.\frac{\mu_{l,\,0}+\mu_{h,\,0}}{m_{\rm PS}^2}\right|
             \left<0|\bar s\gamma_5u|K^+\right>\right|.
\end{equation}
where the normalization is such that $f_\pi\approx130$ MeV, {\it i.e.} larger than
the normalization from our tm$\chi$PT conventions by $\sqrt{2}$.
Unfortunately, the indirect method does not provide easy access to the neutral 
pseudoscalar decay constant, due to mixing with the scalar operator.
The neutral decay constant is not directly accessible in the laboratory due to
the absence of flavor-changing neutral currents in the standard model, but
it is a quantity that appears in the parametrization of some neutral kaon matrix
elements (see Ref.~\cite{dimop} for a very recent example in the context of tmLQCD).
{}From the point of view of our work, 
the comparison of charged and neutral cases would be able to provide 
information about how flavor symmetry breaking in tmLQCD 
affects the structure of mesons.
Note that if we had chosen a different convention in Eq.~\eqref{defdoublets}, 
{\it i.e.} interchanging the role of $c$ and $s$ quarks
as was done in Ref.~\cite{Pena} which focuses on
neutral kaons, then the situation would be reversed: the indirect method 
would have applied to the neutral kaon and not to the charged kaon.

\begin{figure}[tb]
\scalebox{0.38}{\includegraphics*[0cm,11mm][255mm,19cm]{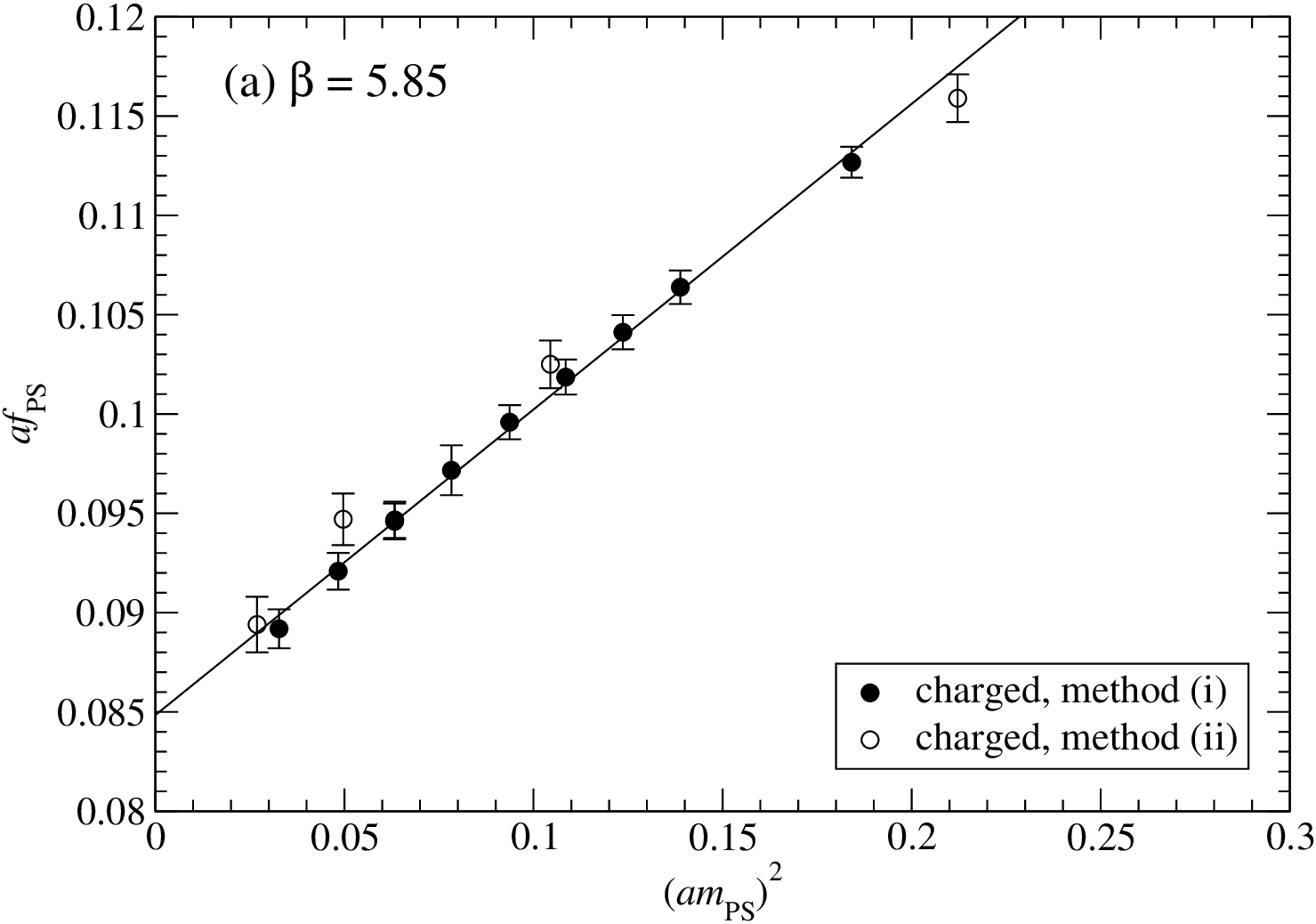}}
\scalebox{0.38}{\includegraphics*[0cm,11mm][255mm,19cm]{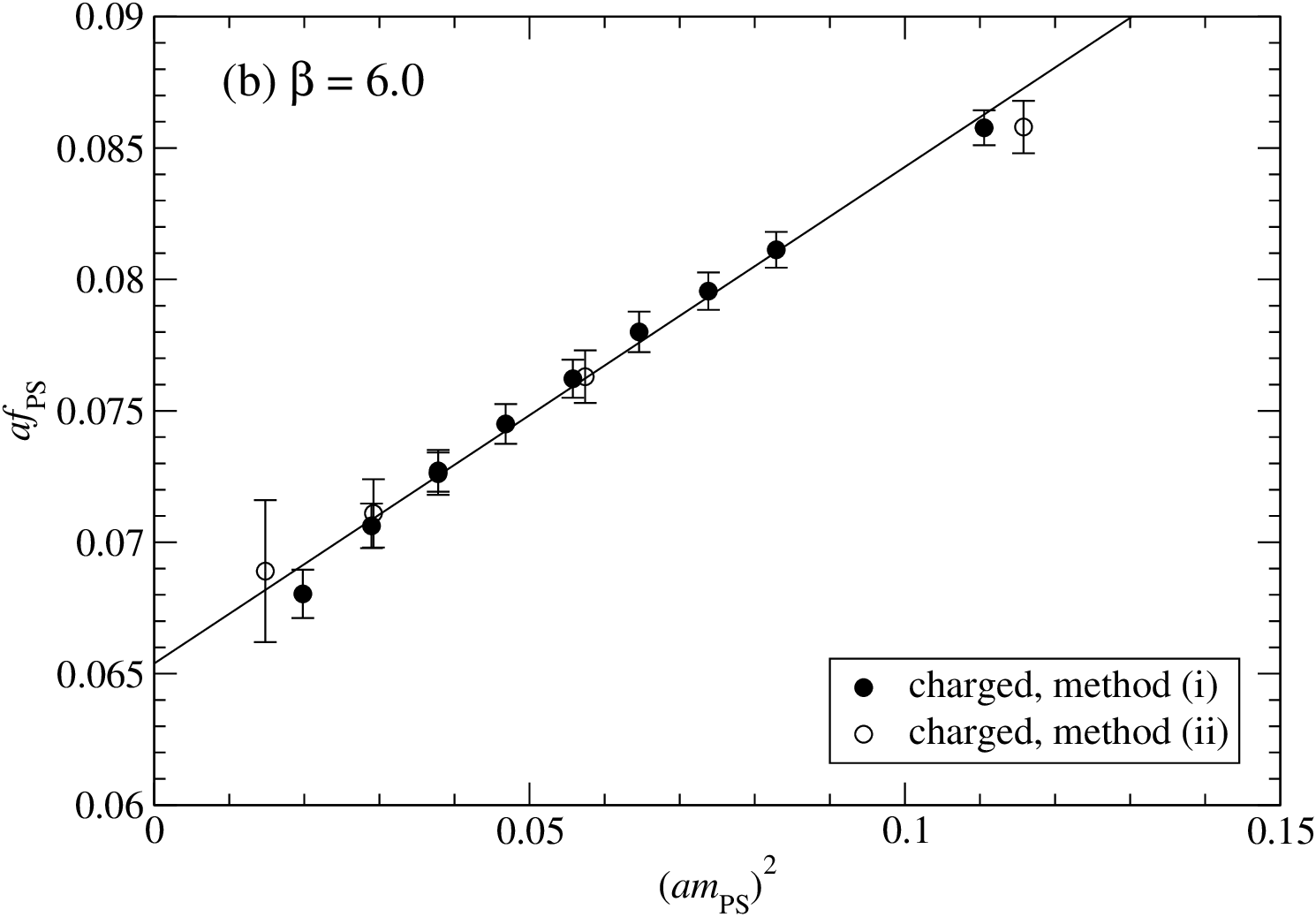}}
\scalebox{0.38}{\includegraphics*[0cm,11mm][255mm,19cm]{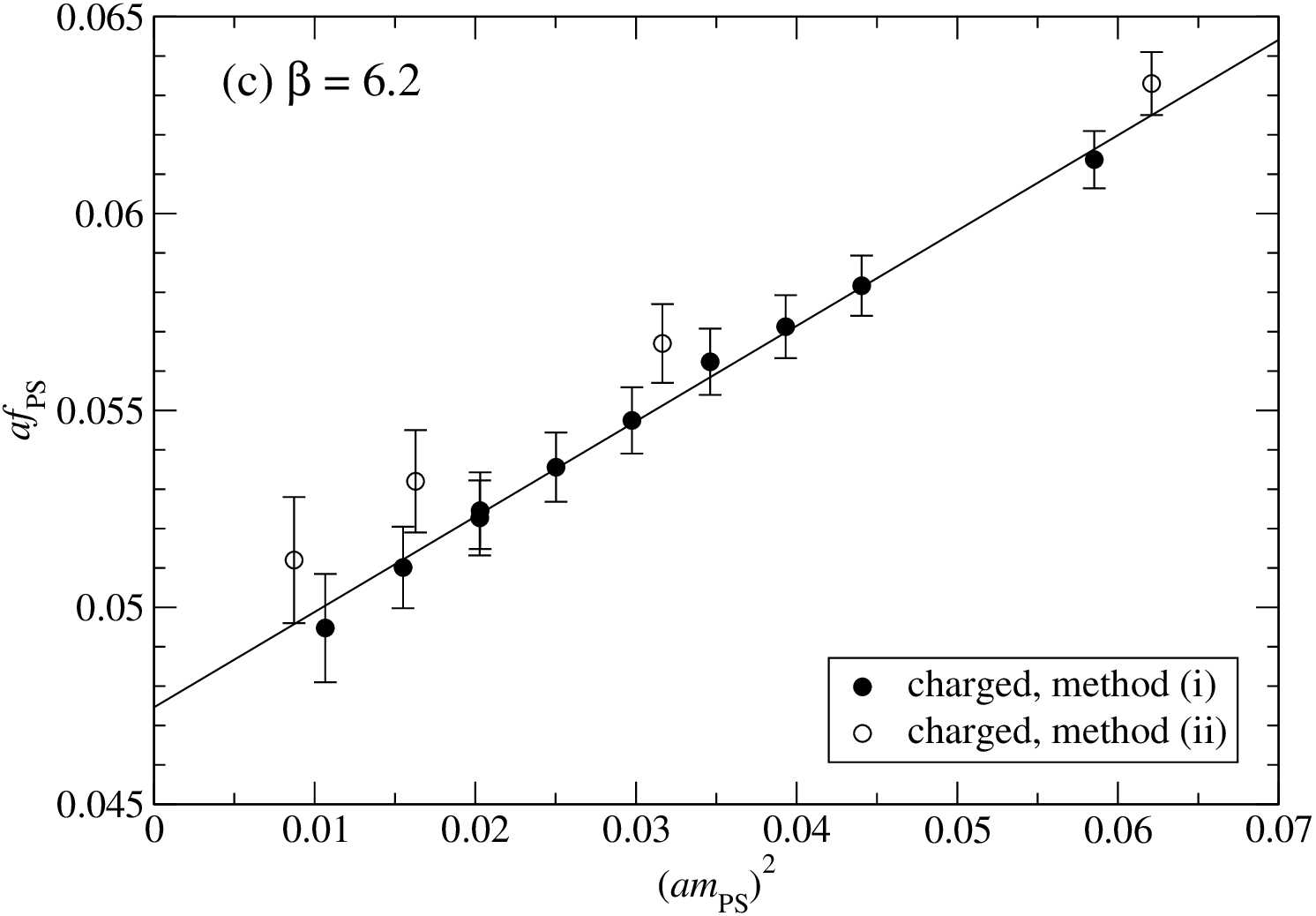}}
\caption{The pseudoscalar meson decay constant as a function of the squared
         charged pseudoscalar meson mass.  Results labelled by method (i)
         are from the present work; results labelled by method (ii) are from
         Ref.~\protect\cite{scaling}.  Straight lines are linear fits
         to the data from method (i).}
\label{fig:fPSvsmPSsq}
\end{figure}

The charged kaon decay constant is plotted in Fig.~\ref{fig:fPSvsmPSsq} where
we continue to define this meson to be
the ground state pseudoscalar meson containing one $s$ (anti)quark from
the heavy doublet and one $u$ (anti)quark from the light doublet.
Central values show a hint of curvature, but within statistical uncertainties
the decay constant is linear in the squared meson mass.  This is consistent
with the tm$\chi$PT expression, i.e. the quenched version of Eq.~\eqref{E:fK}
where the slope has no
logarithmic corrections.  (We have verified that a tm$\chi$PT calculation of
the right-hand side of Eq.~\eqref{indirect} also yields Eq.~\eqref{E:fK}.)
The computations of Ref.~\cite{scaling}, using
method (ii), also appear in Fig.~\ref{fig:fPSvsmPSsq} and are in agreement
with the method (i) results.

\begin{figure}[tb]
\scalebox{0.45}{\includegraphics*[0cm,11mm][255mm,19cm]{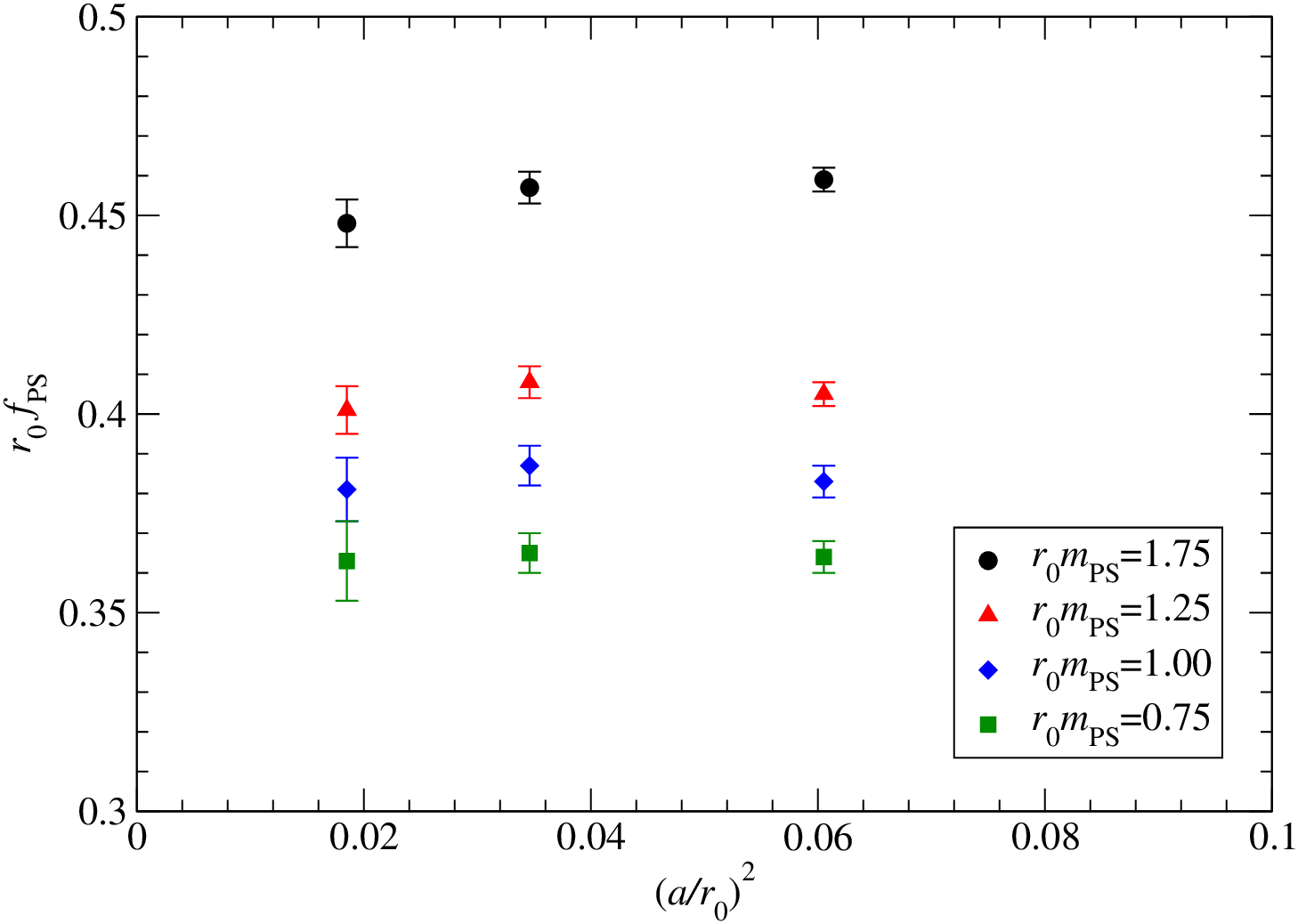}}
\caption{Scaling of the pseudoscalar decay constant for four choices of the
         quark mass.}
\label{fig:fPSscaling}
\end{figure}

As is evident from Fig.~\ref{fig:fPSscaling}, there is no visible dependence
of the decay constant on lattice spacing.  For the kaon, fitting the data
from three $\beta$ values linearly yields
\begin{equation}
r_0m_K = 1.25 \qquad \Rightarrow \qquad f_K = 161 \pm 5 {\rm ~MeV} \,.
\end{equation}
Relying on a linear chiral extrapolation, we similarly obtain
\begin{equation}
f_\pi = 142 \pm 4 {\rm ~MeV} \,.
\end{equation}
The ratio,
\begin{equation}
\frac{f_K}{f_\pi} = 1.136(7) \,,
\end{equation}
agrees nicely with the quenched results from
Ref.~\cite{CPPACS}, though the individual decay constants are somewhat larger.
Using the lattice spacings derived from
Ref.~\cite{Gockeler} or \cite{Dong} would bring us into closer agreement.

\begin{figure}[tb]
\scalebox{0.45}{\includegraphics*[0cm,11mm][255mm,19cm]{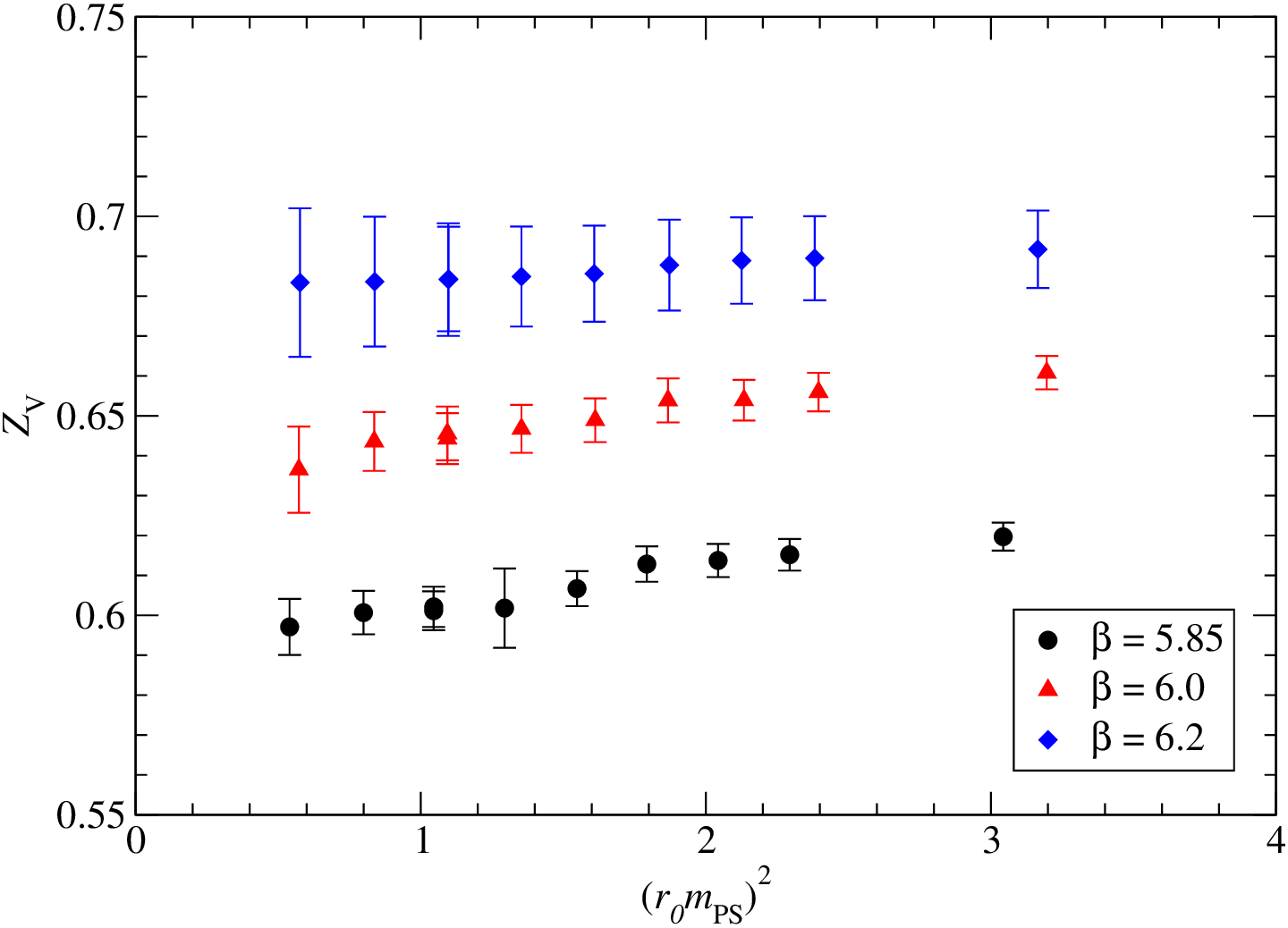}}
\caption{The renormalization factor associated with the pseudoscalar meson
         decay constant.}
\label{fig:ZvsmPSsq}
\end{figure}

There is also a direct method\cite{Bietenholz} for obtaining the decay
constant, though it requires input of a renormalization factor for the
twisted vector current.  Here, we will use the ratio of results from the
direct and indirect methods to determine this renormalization factor.
Figure~\ref{fig:ZvsmPSsq} shows that the renormalization factor is essentially
mass-independent and that it becomes closer to unity as $a\to0$.
The numerical values are comparable to those obtained by the authors of
Ref.~\cite{scaling}, and those authors also note that $Z_V$ is further from
unity in tmLQCD than in both standard and boosted lattice perturbation theory.
(See their Table 7.)

\begin{figure}[tb]
\scalebox{0.38}{\includegraphics*[0cm,11mm][255mm,19cm]{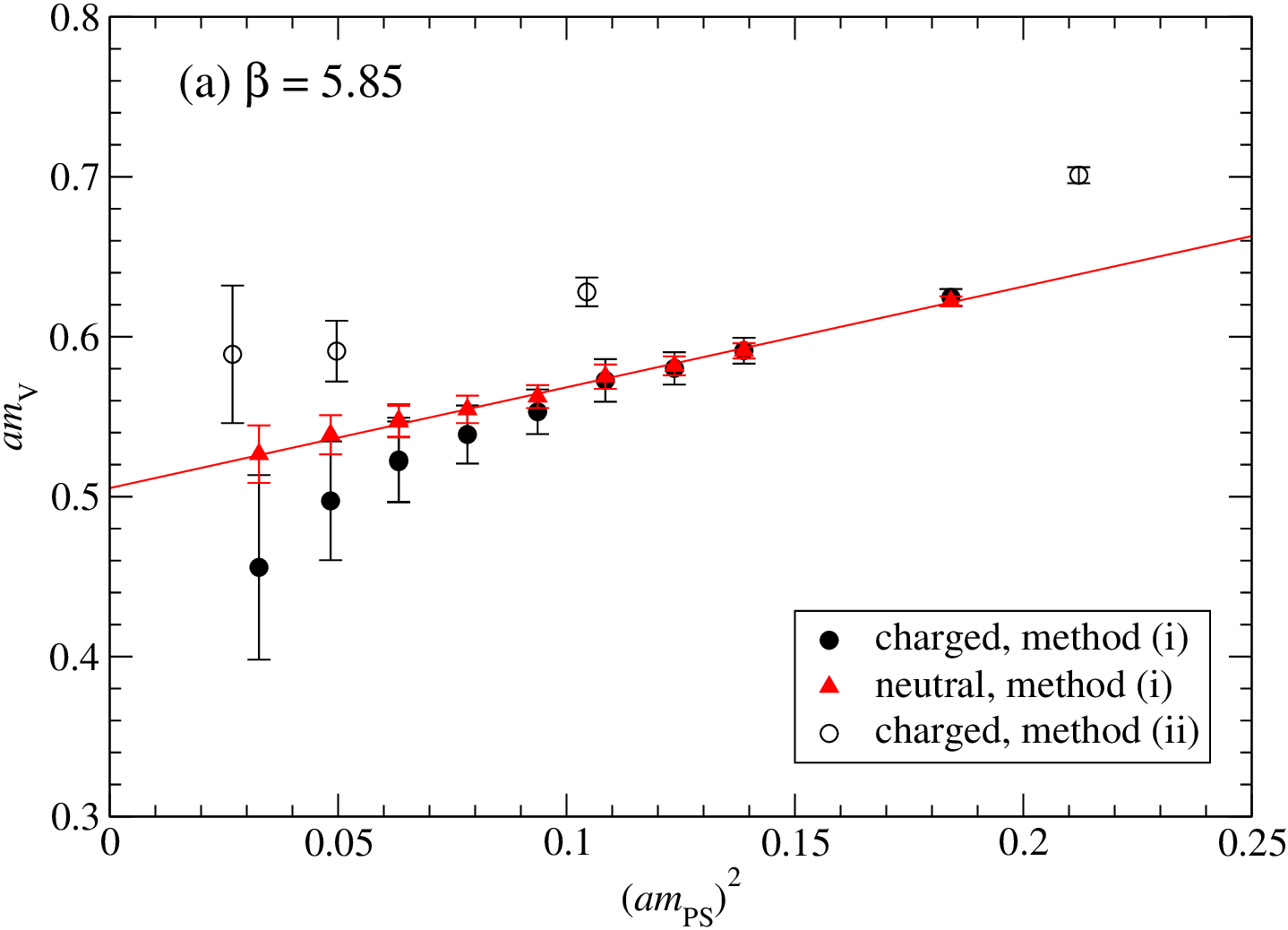}}
\scalebox{0.38}{\includegraphics*[0cm,11mm][255mm,19cm]{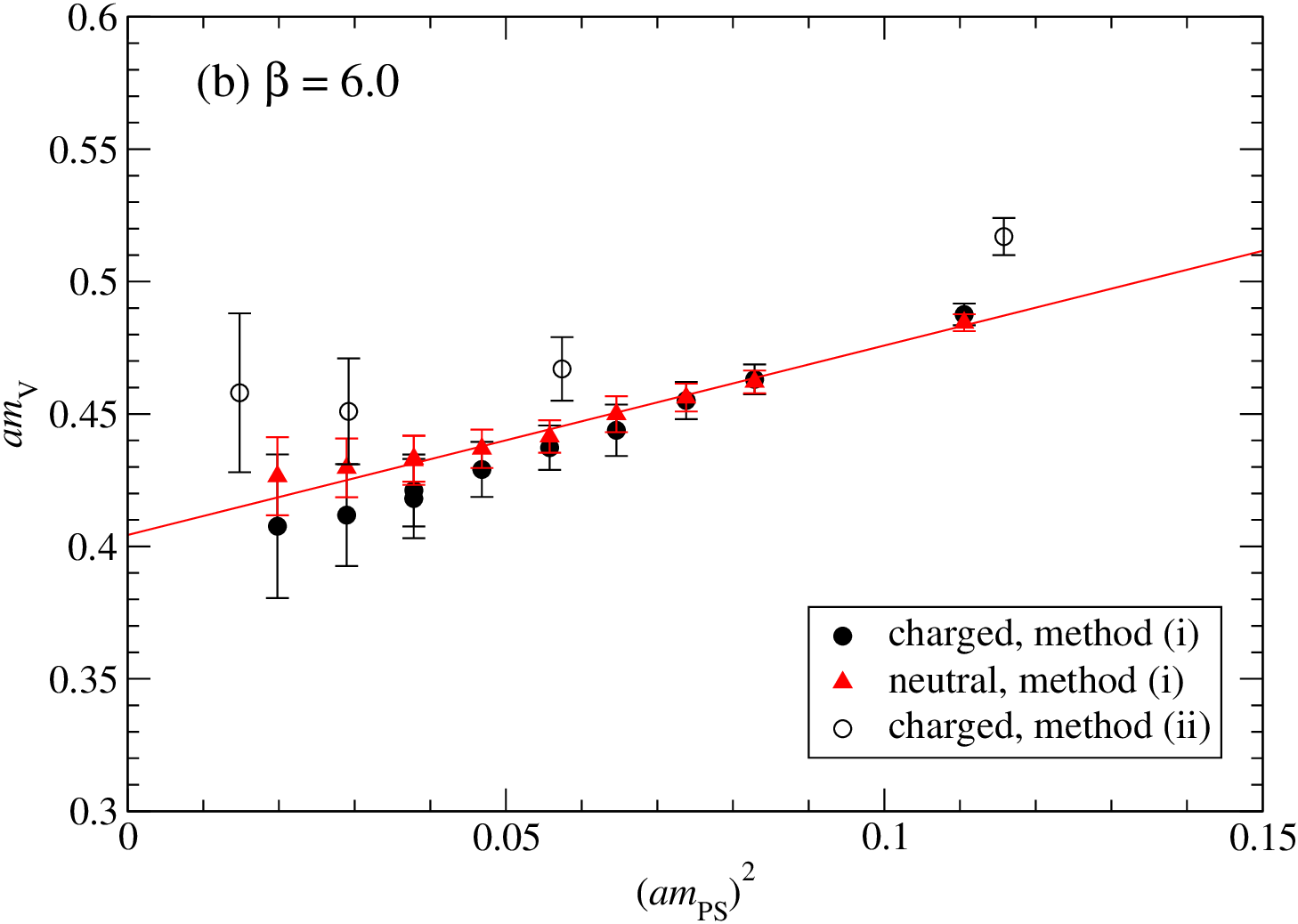}}
\scalebox{0.38}{\includegraphics*[0cm,11mm][255mm,19cm]{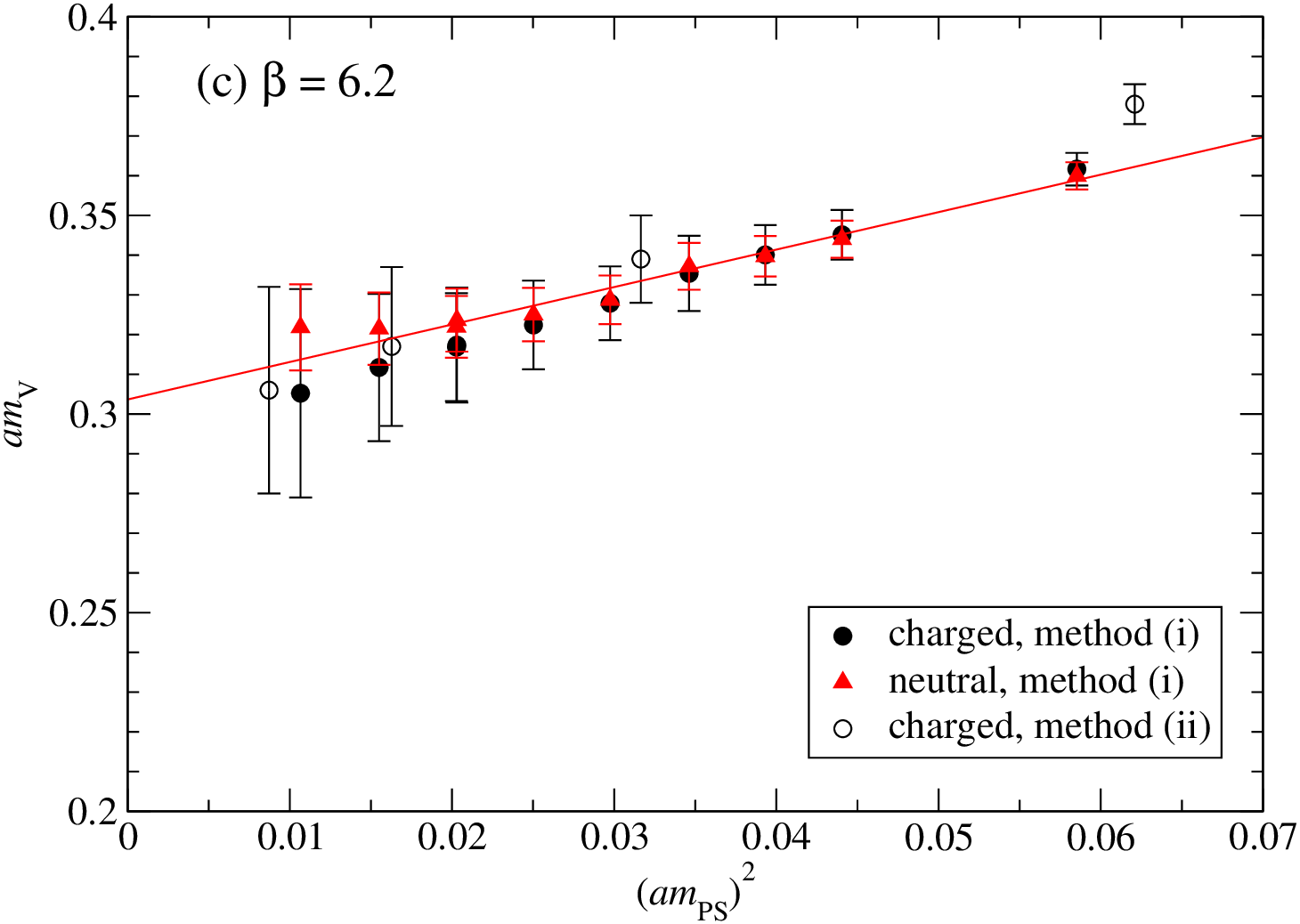}}
\caption{Vector meson mass as a function of the squared charged pseudoscalar
         meson mass.  Results labelled by method (i)
         are from the present work; results labelled by method (ii) are from
         Ref.~\protect\cite{scaling}.  Straight lines are linear fits
         to the neutral data from method (i).}
\label{fig:mVvsmumu}
\end{figure}

Vector meson masses, referred to here as $K^*$ masses
since the strange (anti)quark from the heavy doublet is combined with a
$u$ or $d$ (anti)quark from the light doublet, are
shown in Fig.~\ref{fig:mVvsmumu}, as computed from local operators of the
form $\sum_{k=1}^3\bar\psi\gamma_k\psi$.  Within the statistical uncertainties,
no mass splitting is visible between charged and neutral vector mesons.
Comparison with data from the $\chi$LF collaboration reveals that methods (i)
and (ii) lead to different $K^*$ masses on coarser lattices, and that the
distinction vanishes as $a\to0$.
The sizable uncertainties make chiral extrapolations difficult, particularly
for charged mesons.
Linear fits to the neutral meson masses at each $\beta$ are displayed in
Fig.~\ref{fig:mVvsmumu}.

It is noteworthy that the neutral $K^*$ masses are more precise than the
charged $K^*$, just as the charged pseudoscalar masses are more precise than
the neutral pseudoscalar.  In both cases, the better precision comes in the
channel where the interpolating fields are invariant under twisting.
Moreover, the absence of large cutoff effects for charged pseudoscalars
has been associated with
the existence of an exact lattice axial Ward-Takahashi
identity.\cite{FMPRcutoff}

\begin{figure}[tb]
\scalebox{0.45}{\includegraphics*[0cm,11mm][255mm,19cm]{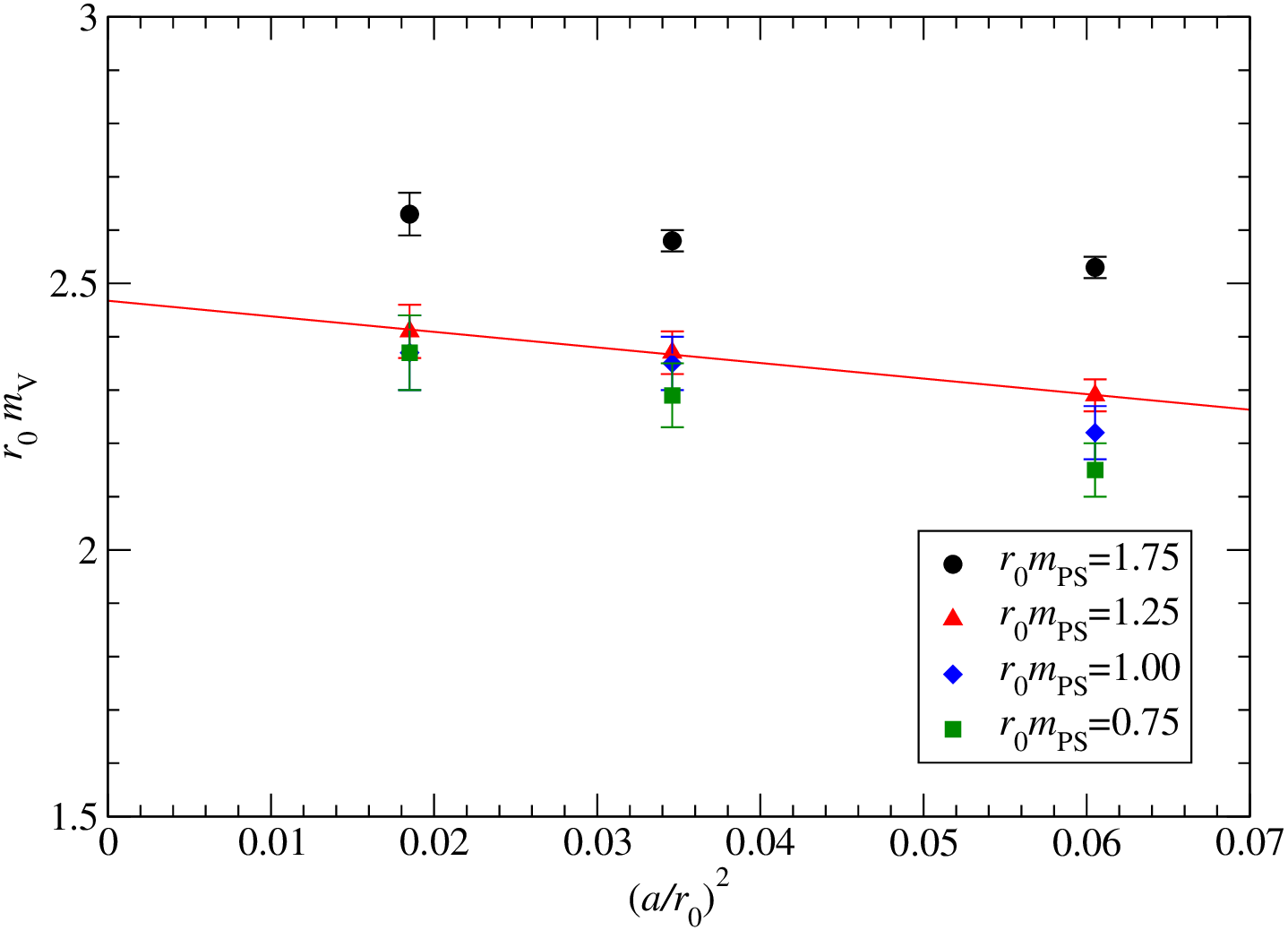}}
\caption{Scaling of the neutral vector meson mass for four choices of the
         quark mass.  The straight line is a linear fit
         to the data having $r_0m_{\rm PS}=1.25$.}
\label{fig:mVscaling}
\end{figure}

Scaling of the neutral vector meson mass with $a^2$ is shown in
Fig.~\ref{fig:mVscaling}.  The nonvanishing dependence on $a^2$ is barely
significant with respect to the uncertainties.
A linear $a^2$ fit to the $r_0m_{\rm PS}=1.25$ data produces
\begin{equation}
m_{K^*} = 970 \pm 20 {\rm ~MeV} \,,
\end{equation}
and a linear $a^2$ fit to the linear chiral extrapolations from
Fig.~\ref{fig:mVvsmumu} yields
\begin{equation}
m_\rho = 916 \pm 20 {\rm ~MeV} \,.
\end{equation}
These quenched values lie above the physical values, but
using the lattice spacings derived from
Ref.~\cite{Gockeler} or \cite{Dong} would bring us closer to experiment.

\section{Scalar meson masses and mixings}\label{sec:scalar}

Our chosen definition of maximal twist, Eq.~\eqref{E:NPwdef}, tunes the
mixing of vector and axial currents for charged mesons, or more precisely, for
mesons built from a quark and antiquark having twist angles of opposite sign.
Charged scalar and pseudoscalar densities do not mix.  Conversely, there is
mixing of neutral scalar and pseudoscalar densities, while neutral vector and
axial currents do not mix.  Our charged vector-axial tuning to maximal twist
can differ from a neutral scalar-pseudoscalar definition, for example by
differing discretization effects.

\begin{figure}[tb]
\scalebox{0.45}{\includegraphics*[0cm,11mm][255mm,19cm]{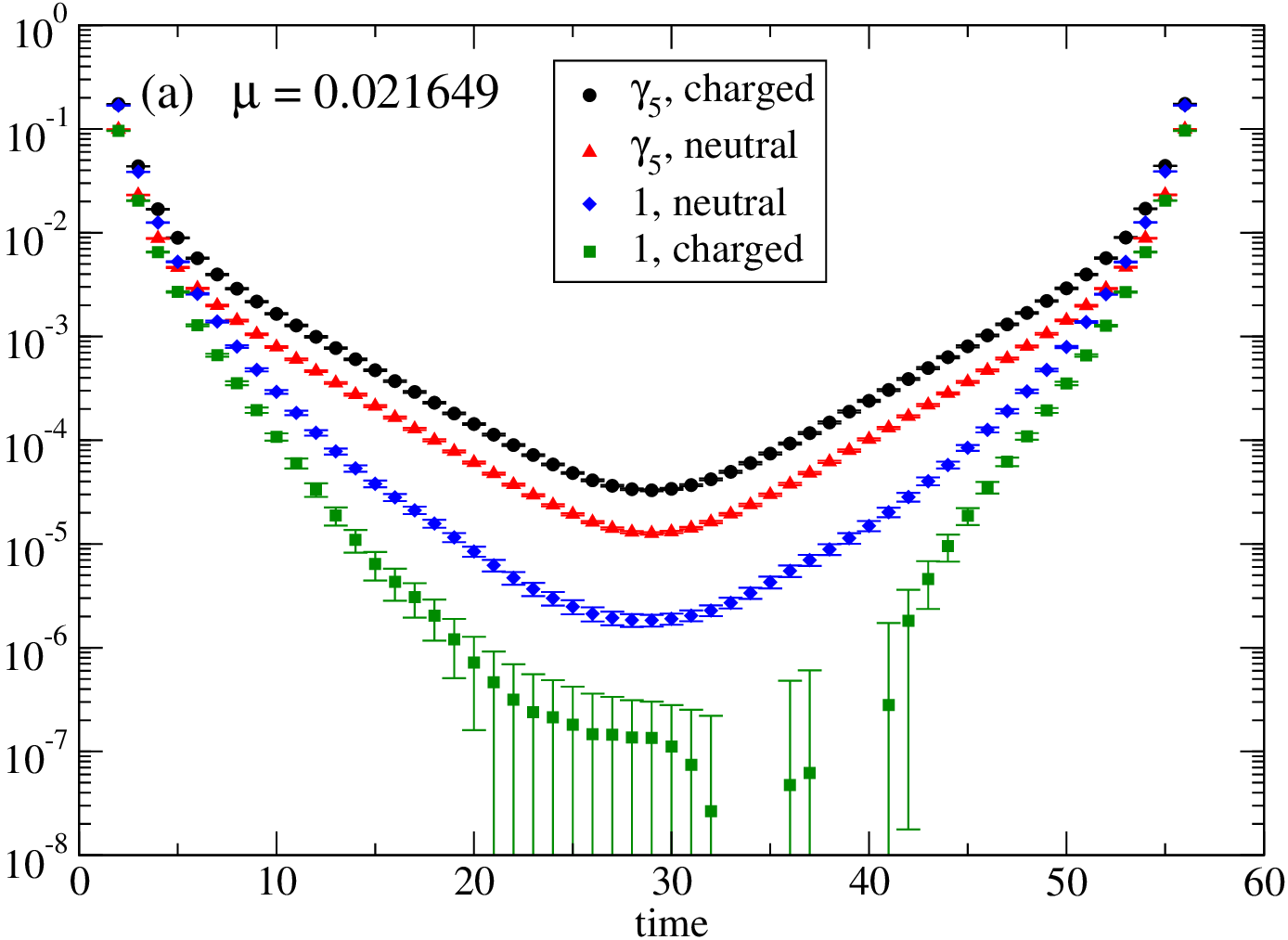}}
\scalebox{0.45}{\includegraphics*[0cm,11mm][255mm,19cm]{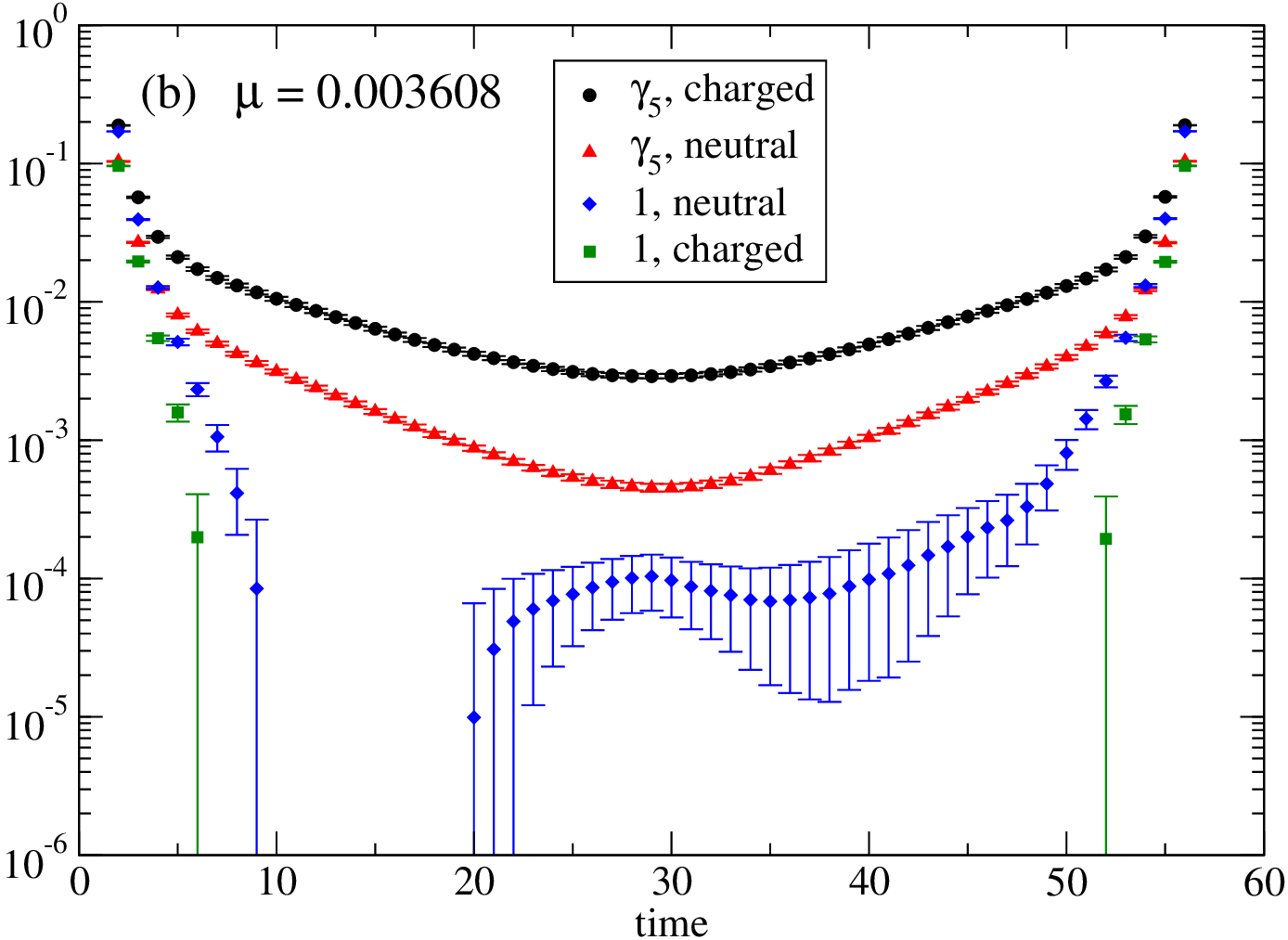}}
\caption{Scalar and pseudoscalar correlation functions for our (a) heaviest and
         (b) lightest quarks at $\beta=6.2$.  The notation $\gamma_5$ and $1$
         refers to the physical basis, defined using method (i).}
\label{fig:b620log}
\end{figure}

Figure \ref{fig:b620log}(a) shows four correlators at $\beta=6.2$ with our
heaviest quark mass: charged and neutral scalar and pseudoscalar
two-point correlators.  The charged correlators cannot mix, and we
see a clear ground state exponential behavior for the pseudoscalar and
scalar mesons, with no contamination between them.  The neutral
pseudoscalar correlator also provides a clear ground state, where the
slope in this log plot is slightly steeper than the charged case, as
expected since we have already established that the neutral meson is heavier
than the charged meson.  The neutral scalar correlator is noticeably different:
it maintains surprisingly small error bars even far from the source, it
displays a kink (change of slope on the log plot) near timesteps 12 and 46,
and it mirrors the neutral pseudoscalar curve between these timesteps.
Apparently the quickly-decaying scalar signal is being
overcome by the pseudoscalar further from the source, {\it i.e.} we are seeing
scalar-pseudoscalar mixing.

\begin{figure}[tb]
\scalebox{0.45}{\includegraphics*[0cm,11mm][255mm,19cm]{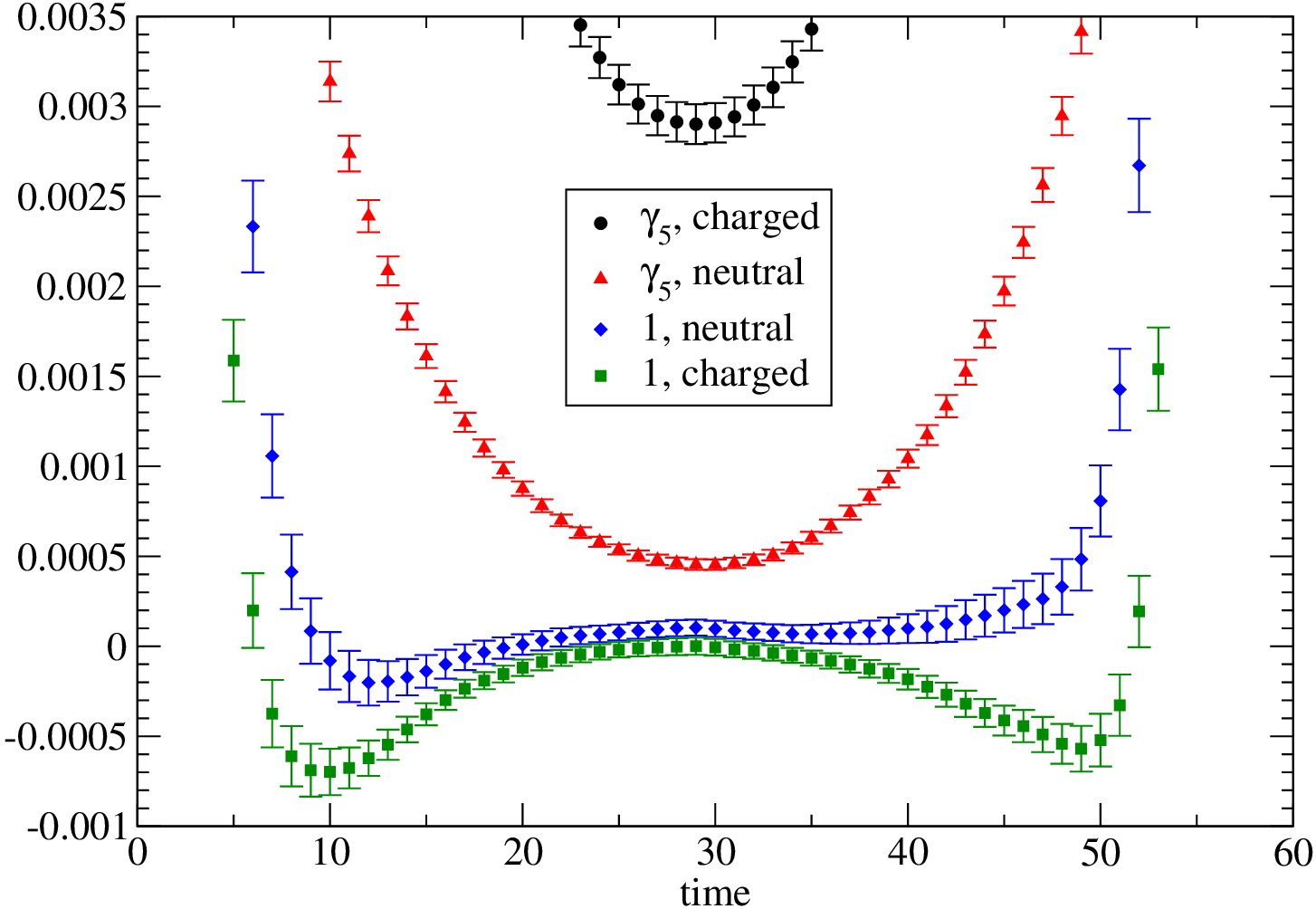}}
\caption{The data from Fig.~\protect\ref{fig:b620log} (b), replotted on a
         linear scale.}
\label{fig:b620lin}
\end{figure}

Figure \ref{fig:b620log}(b) shows the same four correlators but now with our
lightest quark mass.  The effects are now more dramatic, and a new
phenomenon is also observed.  The charged scalar has a brief signal for
the scalar meson near the source, then the correlator becomes negative.
The neutral scalar similarly has a brief signal, then makes a curious
waving shape on the graph.
To understand this, see Fig.~\ref{fig:b620lin} where the data from
Fig.~\ref{fig:b620log}(b) are replotted on a linear scale.
The negative contribution
to the charged scalar correlator is from the two particle state ---
quenched $\eta^\prime$ and kaon --- as discussed in
Refs.~\cite{Bardeen,Kentucky}.
The neutral scalar has this two particle state as well, but
the correlator is deformed because it apparently also has a mixing with
the neutral pseudoscalar.

One could imagine removing the scalar-pseudoscalar mixing by tuning to 
maximal twist directly in this sector, called method (iii) in the notation
of Ref.~\cite{Sharpe}, but mixing could then arise between
vector and axial currents.  An interesting observation, sketched in Fig.~1
of Ref.~\cite{Sharpe}, is that method (iii) is distinct from method (i) but
identical to method (ii) up to $O(a^2)$ corrections.
Could method (ii) be optimal for both the vector-axial and the
scalar-pseudoscalar sectors?

\begin{figure}[tb]
\scalebox{0.45}{\includegraphics*[0cm,11mm][255mm,19cm]{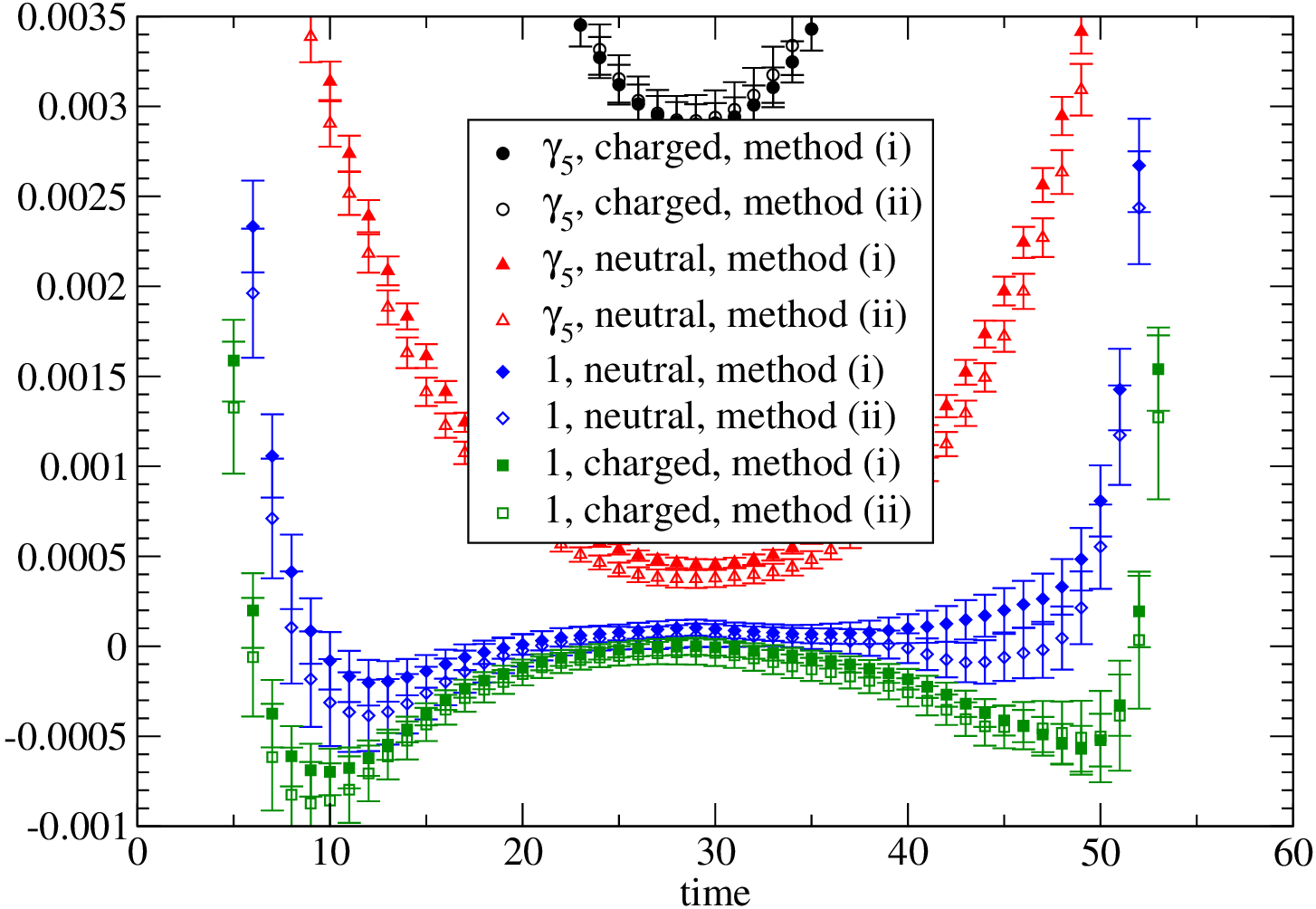}}
\caption{The data from Fig.~\protect\ref{fig:b620lin}, compared to computations
         using method (ii).}
\label{fig:b620linii}
\end{figure}

We have computed 100 quark propagators using method (ii) at $\beta=6.2$
and $\mu_{l,\,0}=0.003608$, allowing a direct comparison to our results from
method
(i).  The method (ii) value of normal quark mass, $m_{l,\,0}=-0.741546$, was
obtained
from Table 3 in Ref.~\cite{flavorbreaking}.  Our findings are displayed in
Fig.~\ref{fig:b620linii}, and we see that the scalar-pseudoscalar mixing is
still apparent, confirming the presence of $O(a^2)$ effects.

\section{Untwisted strange with twisted up and down quarks}\label{sec:Wilson}

If the twist angle, $\omega_h$, of the heavy doublet is set to zero, then
the strange quark becomes a standard Wilson fermion and its partner can be
erased from the action.  Exceptional configurations are typically not a problem
for strange quarks, and $O(a)$ improvement could be accomplished via a
Sheikholeslami-Wohlert term if desired, though we will use the unimproved
Wilson action here.  The $(u,d)$ doublet will be kept at maximal twist.

With this action, the $K^+$ and $\bar K^0$ mesons are exactly
degenerate configuration by configuration since one correlator is the
Hermitian conjugate of the other.  The same is true for $K^-$ and $K^0$.
Furthermore, these two pairs are numerically degenerate in the configuration
average.  This can easily be seen in the free quark limit, since
the twisted mass in one propagator cannot contribute to the correlator
if the other propagator is Wilson, due to the odd number of $\gamma_5$'s.
Similarly, our tm$\chi$PT expression for the mass difference,
Eq.~\eqref{Kmassdiff}, explicitly vanishes when $\omega_h=0$.

\begin{figure}[tb]
\scalebox{0.45}{\includegraphics*[0cm,11mm][255mm,19cm]{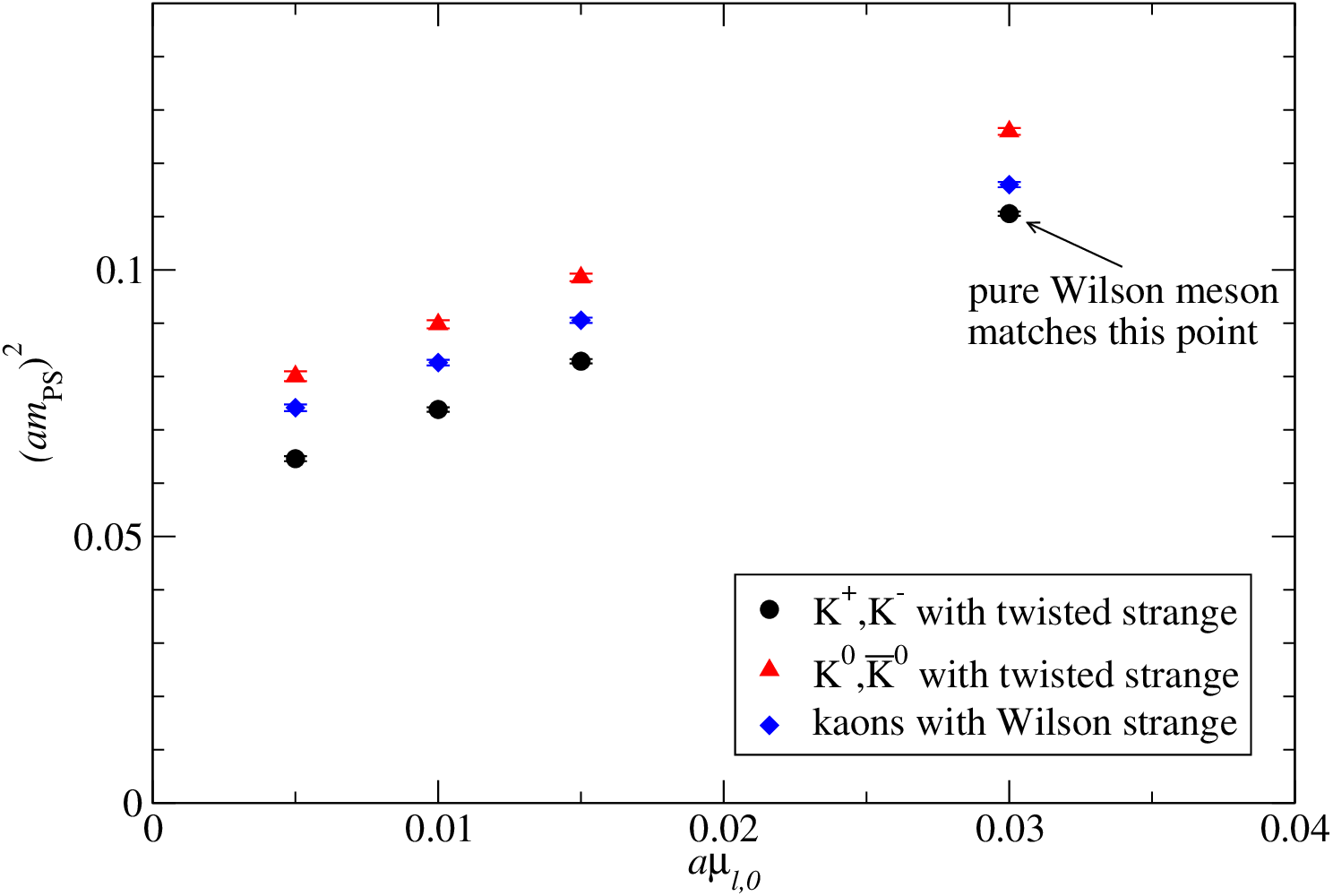}}
\caption{Squared pseudoscalar meson masses with one strange quark/antiquark
         and one light quark/antiquark, plotted as a function of the light
         quark's twisted mass.}
\label{fig:mPSwilsonb600}
\end{figure}

Numerical results at $\beta=6.0$ for kaon masses obtained with a Wilson strange
quark are compared to results with a maximally-twisted strange quark in
Fig.~\ref{fig:mPSwilsonb600}.
For this plot, the twisted strange quark is held fixed at $a\mu_{h,\,0}=0.030$
and the Wilson strange quark's hopping parameter, $\kappa$,
is tuned such that the pseudoscalar mass (obtained from two Wilson propagators
without any twisting)
becomes numerically equal to the charged twisted pseudoscalar mass,
$am_{\rm PS}=0.332(1)$.  The resulting hopping parameter is $\kappa=0.1545$
or equivalently $m_{h,\,0}=-0.7634$.  In both cases, the light quark takes on all
four $(m_{l,\,0},\mu_{l,\,0})$ values from Table~\ref{tab:parameters}.

Fig.~\ref{fig:mPSwilsonb600} shows that the Wilson strange quark leads to 
kaon masses that are numerically between the charged and neutral kaons
with a twisted strange
quark.  All curves are visibly linear in $a\mu_{l,\,0}$, though the Wilson strange
quark theory has a smaller slope than the twisted strange quark theory.
Recall that method (i), used here, itself has a smaller slope than method (ii).

Untwisting the strange quark has other effects besides eliminating the 
mass splitting. When an untwisted quark field is combined with a maximally
twisted one, parity violation induces parity mixing in all (charged and neutral)
channels (this can be inferred from Eq~\eqref{E:JCPhys2TM}). 
This is unlike the situation with complete maximal twisting where 
in some channels the parity violation in the action interchanges parity of the
correlator but does not mix it. With an untwisted strange quark the extraction
of the decay constant becomes a much more difficult problem which is beyond the
scope of the present work.

\section{Summary}\label{sec:outlook}

Twisted mass lattice QCD is a practical method for numerical simulations
involving light quarks.  It has no exceptional configurations, automatic
$O(a)$ improvement, and a corresponding version of chiral perturbation theory.
However, there are issues of parity and flavor symmetry violation effects
at non-zero lattice spacing that have to be understood and dealt with. 

Quarks come in pairs in tmLQCD, so the best way to implement three-flavor
simulations requires thought and exploration.  In this work,
we have considered two doublets at maximal twist, where in our quenched 
simulations the fourth quark is benign. This is in line with the two-doublet
tmLQCD proposed in Ref.~\cite{Pena}.
The chiral perturbation theory is formulated as a natural generalization of
the existing two-flavor formulation, and used to obtain analytic expressions
for masses and decay
constants.  Numerical tmLQCD results for $m_\pi$, $f_\pi$, $m_K$, $f_K$,
$m_\rho$ and $m_{K^*}$ are obtained from four twisted quark masses at each 
of three lattice spacings, and are comparable to previous quenched studies 
with other actions.

Though dynamical simulations were not performed in this study, that is
certainly an ultimate goal for QCD phenomenology.  Dynamical simulations
of the theory with two twisted doublets would mean the fourth quark is no
longer benign.  Identifying it with the physical charm quark requires the
introduction of a mass splitting within the heavy doublet. Progress toward
two-doublet dynamical twisted mass simulations is reviewed in  
Ref.~\cite{Farchioni05}.
Alternatively, one could avoid an active charm quark by using a mixed action
formalism\cite{mixedaction}, for example with twisted $(u,d)$ and untwisted
$s$ quarks
in the sea (recall Sec.~\ref{sec:Wilson}), and Eq.~\eqref{E:action} used for
the valence quarks (so the strange quark's partner is again benign).

One of the significant twist artifacts found in this work is the
mass difference between charged and neutral kaons, which vanishes
in the continuum limit but remains sizable at the lattice spacings studied
here, $0.068$ fm $<a<0.123$ fm.  This splitting depends upon the particular
action that has been chosen; it may be different in other variants of tmLQCD
or in nonquenched simulations.
For example, if only the up and down quarks, not the
strange quark, are twisted, then this large splitting vanishes however at 
the price of a more complicated pattern of parity mixing in the correlators. 
$O(a)$ errors also arise in that scenario, though these could be removed by
the addition of a suitable clover operator.

Another artifact of twisting is the mixing of scalar and pseudoscalar
operators when the standard definitions of maximal twist are employed.
For sufficiently light quarks in the quenched approximation, this can be
studied through the appearance of negative correlators that correspond to the
opening of a quenched $\eta'K$ channel.

Notwithstanding the existence of twisted lattice artifacts, we see value
in the general approach of tmLQCD for applications involving $u$, $d$ and $s$
quarks. There are a number of options for constructing the action including
strange quarks, and with systematic studies such as this one exploring them
we can be hopeful that an optimal approach will be found.

\acknowledgments

RL wishes to thank the TRIUMF Theory Group and the Jefferson Lab Theory
Center for hospitality and support during parts of this research.
RL also thanks Nilmani Mathur for a helpful conversation about scalar mesons
and the quenched $\eta^\prime$.
This work was supported in part by
the Natural Sciences and Engineering Research Council of Canada,
the Canada Foundation for Innovation,
the Canada Research Chairs Program
and the Government of Saskatchewan.

\begin{appendix}
\section{Currents and densities in the twisted basis}\label{app:currents}

With the generators of $SU(2)$ replaced by those of $SU(4)$, the currents and densities 
in the twisted basis of the two-doublet theory are defined in the same way as for the 
one-doublet theory\cite{SW04}. At LO, the currents and densities have the same form, 
\textit{mutatis mutandis}, as in the $SU(2)$ theory. In the physical basis, {\it i.e.}
in terms of the physical variable $\Sigma_{ph}$, they take the forms (with a ``hat''
denoting the physical basis quantity),
\begin{align} \label{E:LOJD}
&V^a_{\mu,\,LO} = \cos(\Delta^a\omega_m)\hat{V}^a_{\mu,\,LO} 
- \eta_{ab}\sin(\Delta^a\omega_m)\hat{A}^b_{\mu,\,LO} \,, \qquad
V^{3,\,8,\,15}_{\mu,\,LO} = \hat{V}^{3,\,8,\,15}_{\mu,\,LO} \,,
\notag \\
&A^a_{\mu,\,LO} = \cos(\Delta^a\omega_m)\hat{A}^a_{\mu,\,LO} 
- \eta_{ab}\sin(\Delta^a\omega_m)\hat{V}^b_{\mu,\,LO} \,, \qquad
A^{3,\,8,\,15}_{\mu,\,LO} = \hat{A}^{3,\,8,\,15}_{\mu,\,LO} \,,
\notag \\
&S^0_{LO} = 
\frac{1}{2}(c_{l,\,m} + c_{h,\,m})\hat{S}^0_{LO} - 2i s_{l,\,m}\hat{P}^3_{LO}
\notag \\
&\qquad\;\;\; + 2(c_{l,\,m} - c_{h,\,m})\left(\frac{1}{\sqrt{3}}\hat{S}^8_{LO} 
+ \frac{1}{\sqrt{6}}\hat{S}^{15}_{LO}\right) - 2i s_{h,\,m} 
\left(\frac{1}{\sqrt{3}}\hat{P}^8_{LO} - \sqrt{\frac{2}{3}}\hat{P}^{15}_{LO}\right) \,,
\notag \\
&S^a_{LO} = \cos(\Sigma^a\omega_m)\hat{S}^a_{LO} 
- i\sin(\Sigma^a\omega_m)\hat{P}^a_{LO} \,, \qquad
P^a_{LO} = \cos(\Sigma^a\omega_m)\hat{P}^a_{LO} 
- i\sin(\Sigma^a\omega_m)\hat{S}^a_{LO} \,, 
\notag \\
&a,\,b \in \mathfrak{K}\backslash\mathfrak{D} \,, \qquad
\mathfrak{K} = \{1,\ldots,15\} \,, \qquad \mathfrak{D} = \{3,\,8,\,15\} \,,
\end{align} 
where $c_{p,\,m} = \cos\omega_{p,\,m}$, $s_{p,\,m} = \sin\omega_{p,\,m}$,
$p = l,\,h$, and we use the notation defined in
Eqs.~(\ref{defapp1}-\ref{defapp3}).
Note that in the $SU(4)$ theory, $P^0_{LO}$ and $S^k_{LO}$, $k\in\mathfrak{K}$,
do not vanish identically in contrast to the $SU(2)$ theory\cite{SW05}.

At NLO, the vector and axial currents are given by
\begin{align}
V^k_\mu &= V^k_{\mu,\,LO}\,(1 + \mathcal{C}) + L_{1,\,2,\,3,\,9}\,\mathrm{terms}
\notag \\
&\quad + L_5\,\frac{1}{2}\mathrm{Tr}\Big[\Big(
[\Lambda_k,\Sigma]_-\,\partial_\mu\Sigma^\dagger 
-\partial_\mu\Sigma\,[\Lambda_k,\Sigma]_-\Big)
(\chi'\Sigma^\dagger + \Sigma\chi'^{\dagger})\Big]
\notag \\
&\quad + \wtil{W}_5\,\frac{1}{2}\mathrm{Tr}\Big[\Big(
[\Lambda_k,\Sigma]_-\,\partial_\mu\Sigma^\dagger 
-\partial_\mu\Sigma\,[\Lambda_k,\Sigma]_-\Big)
(\hat{A}\Sigma^\dagger + \Sigma\hat{A}^{\dagger})\Big] \,, 
\notag \\
A^k_\mu &= A^k_{\mu,\,LO}\,(1 + \mathcal{C}) 
+ \frac{8a W_0}{B_0f^2}W_{10}\,\partial_\mu P^k_{LO}
+ L_{1,\,2,\,3,\,9}\,\mathrm{terms} 
\notag \\
&\quad - L_5\,\frac{1}{2}\mathrm{Tr}\Big[\Big(
[\Lambda_k,\Sigma]_+\,\partial_\mu\Sigma^\dagger 
-\partial_\mu\Sigma\,[\Lambda_k,\Sigma]_+\Big)
(\chi'\Sigma^\dagger + \Sigma\chi'^{\dagger})\Big] 
\notag \\
&\quad - \wtil{W}_5\,\frac{1}{2}\mathrm{Tr}\Big[\Big(
[\Lambda_k,\Sigma]_+\,\partial_\mu\Sigma^\dagger 
-\partial_\mu\Sigma\,[\Lambda_k,\Sigma]_+\Big)
(\hat{A}\Sigma^\dagger + \Sigma\hat{A}^{\dagger})\Big] \,,
\end{align}
where $k\in\mathfrak{K}$, and
\begin{align} \label{E:Cdef}
\mathcal{C} &= \frac{4L_4}{f^2}\,\mathrm{Tr}
[\chi'^{\dagger}\Sigma+\Sigma^\dagger\chi']
+\frac{4\wtil{W}_4}{f^2}\mathrm{Tr}(\hat{A}^\dagger\Sigma + \Sigma^\dagger\hat{A})
%- \frac{32a W_0}{2B_0f^4}\wtil{W}_4S_{LO}^0 \,.
\end{align}
We do not give the form of the $L_{1,\,2,\,3,\,9}$ terms since each has the
same form as in the continuum $SU(2)$ theory\cite{GL84}. 

Dropping terms proportional to the scalar and pseudoscalar sources, which give rise 
only to contact terms in correlation functions, the scalar and pseudoscalar densities 
at NLO are given by 
\begin{align}
S^k &= S^k_{LO}\,(1 + \mathcal{D}_1) + P^k_{LO}\,\mathcal{D}_2
+ L_5\,B_0\mathrm{Tr}\big[\partial_\mu\Sigma\partial_\mu\Sigma^\dagger
(\Lambda_k\Sigma^\dagger + \Sigma\Lambda_k)\big]
\notag \\
&\quad - L_8\,2B_0\mathrm{Tr}\big[(\Lambda_k\Sigma^\dagger + \Sigma\Lambda_k)
(\chi'\Sigma^\dagger + \Sigma\chi'^{\dagger})\big]
- \wtil{W}_8\,B_0\mathrm{Tr}\big[(\Lambda_k\Sigma^\dagger + \Sigma\Lambda_k)
(\hat{A}\Sigma^\dagger + \Sigma\hat{A}^\dagger)\big] \,,
\notag \\
P^k &= P^k_{LO}\,(1 + \mathcal{D}_1) + S^k_{LO}\,\mathcal{D}_2
+ L_5\,B_0\mathrm{Tr}\big[\partial_\mu\Sigma\partial_\mu\Sigma^\dagger
(\Lambda_k\Sigma^\dagger - \Sigma\Lambda_k)\big]
\notag \\
&\quad - L_8\,2B_0\mathrm{Tr}\big[(\Lambda_k\Sigma^\dagger - \Sigma\Lambda_k)
(\chi'\Sigma^\dagger + \Sigma\chi'^{\dagger})\big]
- \wtil{W}_8\,B_0\mathrm{Tr}\big[(\Lambda_k\Sigma^\dagger - \Sigma\Lambda_k)
(\hat{A}\Sigma^\dagger + \Sigma\hat{A}^\dagger)\big] \,,
\notag \\
&\quad + 4i H_2 B_0^2\mathrm{Tr}(\Lambda_k\bsym{\mu}) \,,
\end{align}
where $k\in\mathfrak{K}$, and
\begin{align}
\mathcal{D}_1 &=  
-\frac{4L_4}{f^2}\,\mathrm{Tr}[D_\mu\Sigma D_\mu\Sigma^\dagger]
+\frac{8L_6}{f^2}\,\mathrm{Tr}(\chi'^{\dagger}\Sigma +\Sigma^\dagger\chi) 
+\frac{4\wtil{W}_6}{f^2}\mathrm{Tr}(\hat{A}^\dagger\Sigma + \Sigma^\dagger\hat{A}) \,, 
\notag \\
\mathcal{D}_2 &= 
-\frac{8L_7}{f^2}\mathrm{Tr}(\chi'^{\dagger}\Sigma - \Sigma^\dagger\chi')
-\frac{4\wtil{W}_7}{f^2}\mathrm{Tr}(\hat{A}^\dagger\Sigma - \Sigma^\dagger\hat{A}) \,.
\end{align}

To write the NLO currents and densities in the physical basis, we need the results
\begin{align}
&\mathrm{Tr}(D_\mu^\dagger\Sigma D_\mu\Sigma^\dagger) =
\mathrm{Tr}(D_\mu^\dagger\Sigma_{ph}D_\mu\Sigma_{ph}^\dagger) \,, 
\notag \\
&\mathrm{Tr}(\chi'^{\dagger}\Sigma\pm\Sigma^\dagger\chi) =
\begin{cases}
-\frac{4\hat{M}'}{2B_0 f^2}\hat{S}^0_{LO} 
-\frac{4\Delta M'}{B_0 f^2}
(\frac{1}{\sqrt{3}}\hat{S}^8_{LO} +\frac{1}{\sqrt{6}}\hat{S}^{15}_{LO}) 
\quad (+\;\mathrm{sign}) \\
\quad\!\frac{4\hat{M}'}{2B_0 f^2}\hat{P}^0_{LO} 
+\frac{4\Delta M'}{B_0 f^2}
(\frac{1}{\sqrt{3}}\hat{P}^8_{LO} +\frac{1}{\sqrt{6}}\hat{P}^{15}_{LO})
\quad (-\;\mathrm{sign})
\end{cases}
+ O(M_p'\epsilon_p) \,,
\notag \\
&\mathrm{Tr}(\hat{A}^\dagger\Sigma\pm\Sigma^\dagger\hat{A}) =
\begin{cases}
-\frac{8W_0 a}{2B_0 f^2}\hat{S}^0_{LO} \quad (+\;\mathrm{sign}) \\
\quad\!\frac{8W_0 a}{2B_0 f^2}\hat{P}^0_{LO} \quad (-\;\mathrm{sign})
\end{cases}
+ O(a\epsilon_p) \,, \qquad p = l,\,h \,,
\end{align}
where 
\begin{equation}
\hat{M}' \equiv (M_l' + M_h')/2 \,, \qquad \Delta M' \equiv M_l' - M_h' \,,
\end{equation}
which allow us to express $\mathcal{C}$, $\mathcal{D}_1$, and $\mathcal{D}_2^\pm$ in 
terms of the physical fields. 

Next we write the $L_5$, $L_8$, $\wtil{W}_5$, $\wtil{W}_8$, and $H_2$ terms 
in the physical basis. For the $L_5$ terms, we need the results
\begin{align}
&\pm\frac{1}{2}\mathrm{Tr}\Big[\Big(
[\Lambda_a,\Sigma]_\mp\,\partial_\mu\Sigma^\dagger 
-\partial_\mu\Sigma\,[\Lambda_a,\Sigma]_\mp\Big)
(\chi'\Sigma^\dagger + \Sigma\chi'^{\dagger})\Big] \notag \\
&\qquad =
\begin{cases}
\cos(\Delta^a\omega_m)\hat{V}^a_{L_5} - \eta_{ab}\sin(\Delta^a\omega_m)\hat{A}^b_{L_5} 
\quad (\mathrm{upper\;sign}) \\
\cos(\Delta^a\omega_m)\hat{A}^a_{L_5} - \eta_{ab}\sin(\Delta^a\omega_m)\hat{V}^b_{L_5} 
\quad (\mathrm{lower\;sign})
\end{cases}, 
\notag \\
&\mathrm{Tr}\big[\partial_\mu\Sigma\partial_\mu\Sigma^\dagger
(\Lambda_a\Sigma^\dagger\pm\Sigma\Lambda_a)\big] =
\begin{cases}
\cos(\Sigma^a\omega_m)\hat{S}^a_{L_5} - i\sin(\Sigma^a\omega_m)\hat{P}^a_{L_5}
\quad (+\;\mathrm{sign}) \\
\cos(\Sigma^a\omega_m)\hat{P}^a_{L_5} - i\sin(\Sigma^a\omega_m)\hat{S}^a_{L_5}
\quad (-\;\mathrm{sign})
\end{cases},
\end{align}
where $a,\,b\in\mathfrak{K}\backslash\mathfrak{D}$, and
\begin{align}
\hat{V}^a_{L_5} &= \frac{1}{2}\left\langle\Lambda_a\Big(
\big[\Sigma_{ph}\partial\Sigma^\dagger_{ph},
\chi_{tw}'\Sigma^\dagger + \Sigma_{ph}\chi_{tw}'^{\dagger}\big]_{+}
+ (\Sigma\leftrightarrow\Sigma^\dagger,\,\chi'\leftrightarrow\chi'^{\dagger})
\Big)\right\rangle \,, \qquad 
\chi_{tw}' \equiv \xi_m^\dagger\chi'\xi_m^\dagger
\notag \\
\hat{A}^a_{L_5} &= \frac{1}{2}\left\langle\Lambda_a\Big(
\big[\Sigma_{ph}^\dagger\partial\Sigma_{ph},
\chi_{tw}'^{\dagger}\Sigma + \Sigma_{ph}^\dagger\chi_{tw}'\big]_{+}
- (\Sigma\leftrightarrow\Sigma^\dagger,\,\chi'\leftrightarrow\chi'^{\dagger})
\Big)\right\rangle \,,
\notag \\
\hat{S}^a_{L_5} &= -\left\langle\Lambda_a\big(
\partial_\mu\Sigma_{ph}\Sigma_{ph}^\dagger\partial_\mu\Sigma_{ph} + \mathrm{h.c.}
\big)\right\rangle \,, \qquad
\hat{P}^a_{L_5} = \left\langle\Lambda_a\big(
\partial_\mu\Sigma_{ph}\Sigma_{ph}^\dagger\partial_\mu\Sigma_{ph} - \mathrm{h.c.}
\big)\right\rangle \,,
\end{align}
and for the $L_8$ terms we need the results
\begin{align}
&-\mathrm{Tr}\big[(\Lambda_a\Sigma^\dagger\pm\Sigma\Lambda_a)
(\chi'\Sigma^\dagger + \Sigma\chi'^{\dagger})\big] = 
\begin{cases}
\cos(\Sigma^a\omega_m)\hat{S}^a_{L_8} - i\sin(\Sigma^a\omega_m)\hat{P}^a_{L_8}
\quad (+\;\mathrm{sign}) \\
\cos(\Sigma^a\omega_m)\hat{P}^a_{L_8} - i\sin(\Sigma^a\omega_m)\hat{S}^a_{L_8}
\quad (-\;\mathrm{sign})
\end{cases},
\notag \\
&\hat{S}^a_{L_8} = -\left\langle\Lambda_a\big(
\Sigma_{ph}\chi_{tw}'^{\dagger}\Sigma_{ph} + \mathrm{h.c.}\big)\right\rangle \,,
\qquad
\hat{P}^a_{L_8} = \left\langle\Lambda_a\big(
\Sigma_{ph}\chi_{tw}'^{\dagger}\Sigma_{ph} - \mathrm{h.c.}\big)\right\rangle \,,
\qquad a \in \mathfrak{K}\backslash\mathfrak{D} \,.
\end{align}
By replacing $\chi'$ with $\hat{A}$, and $L_{5,\,8}$ with $\wtil{W}_{5,\,8}$, the 
$\wtil{W}_5$ and $\wtil{W}_8$ terms can be expressed in the physical basis using the 
same results above for the $L_5$ and $L_8$ terms.

Lastly, the $H_2$ term contributes only in the flavor-diagonal case, {\it i.e.} when 
the flavor index $k = 3,\,8,\,15$. Since we will not be using the flavor-diagonal
currents and densities, we do not give results for the flavor-diagonal cases here.

To conclude this appendix, we provide the explicit expression for the axial
current in the physics basis at NLO, using the twist angles determined in
Sec.~\ref{subsec:twangle}:
\begin{align}
\hat{A}^a_\mu &= \hat{A}^a_{\mu,\,LO} - \frac{8W_0 a}{f^2}W_{10}
\cos(\Sigma^a\omega_m)\sin(\Delta^a\omega_m)\eta_{ab}\hat{V}^b_{\mu,\,LO}
(1 + \mathcal{C}) + L_{1,\,2,\,3,\,9}\,\mathrm{terms} \notag \\
&\quad +\frac{8W_0 a}{B_0 f^2}W_{10}\cos(\Delta^a\omega_m)
\Big[\cos(\Sigma^a\omega_m)\partial_\mu\hat{P}^a_{LO} - 
i\sin(\Sigma^a\omega_m)\partial_\mu\hat{S}^a_{LO}\Big] \notag \\
&\quad +L_5\bigg[\hat{A}^a_{L_5} - \frac{8W_0 a}{f^2}W_{10}
\cos(\Sigma^a\omega_m)\sin(\Delta^a\omega_m)\eta_{ab}\hat{V}^b_{L_5}\bigg]
+\big[L_5 \leftrightarrow \wtil{W}_5\big] \,.
\end{align}
The term $\mathcal{C}$ is the same as in Eq.~\eqref{E:Cdef}, but now given in
the physical basis.
\end{appendix}

\newpage

% table 1
\begin{table}
\caption{The parameters used for simulations in this work.
Lattice spacings are taken from Ref.~\protect\cite{Jansen1}.
Each $(am_{p,\,0},a\mu_{p,\,0})$ pair is the result of tuning
to maximal twist with method (i)
as discussed in Sec.~\protect\ref{sec:details}.
The subscript $p=l,h$ is used in the text to distinguish the
``light'' quark doublet from the ``heavy'' quark doublet, but for
purposes of numerical tuning in this table there is no
distinction. The twist angle was obtained from Eq.~(\protect\ref{defangle}).
}\label{tab:parameters}
\begin{tabular}{ccccclr@{$\pm$}l}
$\beta$~~ & ~~$a$ [fm]~~ & \#sites & ~~\#configurations
 & ~~~~$am_{p,\,0}$~~~~ & ~~$a\mu_{p,\,0}$
& \multicolumn{2}{c}{twist angle (degrees)} \\
\hline
5.85 & 0.123 & $20^3\times40$ & 600 & -0.8965 & 0.0376 & ~~~~~90.0 & 0.3 \\
     &       &                &     & -0.9071 & 0.0188 & 90.2 & 0.6 \\
     &       &                &     & -0.9110 & 0.01252 & 90.6 & 0.8 \\
     &       &                &     & -0.9150 & 0.00627 & 90.6 & 1.6 \\
\hline
6.0  & 0.093 & $20^3\times48$ & 600 & -0.8110 & 0.030 & 90.4 & 0.4 \\
     &       &                &     & -0.8170 & 0.015 & 91.0 & 0.7 \\
     &       &                &     & -0.8195 & 0.010 & 92.5 & 1.0 \\
     &       &                &     & -0.8210 & 0.005 & 95.5 & 2.1 \\
\hline
6.2  & 0.068 & $28^3\times56$ & 200 & -0.7337 & 0.021649 & 89.1 & 0.8 \\
     &       &                &     & -0.7367 & 0.010825 & 87.3 & 1.8 \\
     &       &                &     & -0.7378 & 0.007216 & 86.3 & 2.8 \\
     &       &                &     & -0.7389 & 0.003608 & 86.4 & 4.5
\end{tabular}
\end{table}

% table 2
\begin{table}
\caption{Numerical values from our simulations.  These are also shown
         graphically in the figures.  The rows with superscripts $a$ and $b$
         refer to ($a\mu_{h,0}$,$a\mu_{l,0}$)=(0.015,0.005) and (0.010,0.010)
         respectively.}\label{tab:allnumbers}
\begin{tabular}{llllllll}
$\beta$ & $a\mu_{l,\,0}+a\mu_{h,\,0}$ & \multicolumn{2}{c}{$(am_{\rm PS})^2$} &
$af_{\rm PS}$ & $Z_V$ & \multicolumn{2}{c}{$am_V$} \\
\cline{3-4}\cline{7-8}
 & & charged & neutral & & & charged & neutral \\
\hline
5.85 & 0.0752 & 0.1841(4) & 0.2262(12) & 0.1127(8) & 0.620(4) & 0.625(5) & 0.622(3) \\
     & 0.0564 & 0.1388(4) & 0.1827(14) & 0.1064(8) & 0.615(4) & 0.591(8) & 0.591(5) \\
     & 0.05012 & 0.1236(4) & 0.1683(15) & 0.1041(9) & 0.614(4) & 0.580(10) & 0.582(6) \\
     & 0.04387 & 0.1085(4) & 0.1536(19) & 0.1019(9) & 0.613(4) & 0.573(13) & 0.575(8) \\
     & 0.0376 & 0.0937(3) & 0.1402(15) & 0.0996(9) & 0.607(4) & 0.553(14) & 0.563(7) \\
     & 0.03132 & 0.0784(10) & 0.1259(16) & 0.0972(13) & 0.602(10) & 0.539(18) & 0.555(9) \\
     & 0.02507 & 0.0633(3) & 0.1112(19) & 0.0947(9) & 0.602(5) & 0.523(26) & 0.548(10) \\
     & 0.02504 & 0.0633(3) & 0.1117(17) & 0.0946(9) & 0.601(5) & 0.522(25) & 0.547(10) \\
     & 0.01879 & 0.0484(2) & 0.0969(22) & 0.0921(9) & 0.601(5) & 0.497(37) & 0.539(12) \\
     & 0.01254 & 0.0327(2) & 0.0824(33) & 0.0892(10) & 0.597(7) & 0.456(58) & 0.527(18) \\
\hline
6.0  & 0.060 & 0.1106(4) & 0.1260(6) & 0.0858(7) & 0.661(4) & 0.488(4) & 0.484(3) \\
     & 0.045 & 0.0829(4) & 0.0986(7) & 0.0811(7) & 0.656(5) & 0.463(6) & 0.462(4) \\
     & 0.040 & 0.0738(4) & 0.0898(8) & 0.0796(7) & 0.654(5) & 0.455(7) & 0.456(5) \\
     & 0.035 & 0.0646(5) & 0.0801(9) & 0.0780(8) & 0.654(6) & 0.444(10) & 0.450(7) \\
     & 0.030 & 0.0558(4) & 0.0724(7) & 0.0762(7) & 0.649(5) & 0.437(8) & 0.441(6) \\
     & 0.025 & 0.0468(4) & 0.0641(8) & 0.0745(8) & 0.647(6) & 0.429(10) & 0.437(7) \\
     & 0.020$^a$ & 0.0378(4) & 0.0550(10) & 0.0726(8) & 0.646(7) & 0.418(15) & 0.432(9) \\
     & 0.020$^b$ & 0.0378(4) & 0.0559(9) & 0.0727(8) & 0.644(6) & 0.421(14) & 0.433(9) \\
     & 0.015 & 0.0290(3) & 0.0471(12) & 0.0706(8) & 0.644(7) & 0.412(19) & 0.430(11) \\
     & 0.010 & 0.0198(3) & 0.0383(18) & 0.0680(9) & 0.637(11) & 0.407(27) & 0.426(15) \\
\hline
6.2  & 0.043298 & 0.0585(4) & 0.0640(6) & 0.0614(7) & 0.692(10) & 0.362(4) & 0.360(3) \\
     & 0.032474 & 0.0441(4) & 0.0497(6) & 0.0582(8) & 0.689(11) & 0.345(6) & 0.344(5) \\
     & 0.028865 & 0.0393(4) & 0.0451(7) & 0.0571(8) & 0.689(11) & 0.340(8) & 0.340(5) \\
     & 0.025257 & 0.0346(5) & 0.0406(7) & 0.0562(8) & 0.688(11) & 0.335(9) & 0.337(6) \\
     & 0.02165 & 0.0298(4) & 0.0358(7) & 0.0547(8) & 0.686(12) & 0.328(9) & 0.329(6) \\
     & 0.018041 & 0.0250(4) & 0.0313(7) & 0.0536(9) & 0.685(13) & 0.322(11) & 0.325(7) \\
     & 0.014433 & 0.0203(4) & 0.0269(8) & 0.0525(10) & 0.684(13) & 0.317(14) & 0.324(8) \\
     & 0.014432 & 0.0203(4) & 0.0268(8) & 0.0523(10) & 0.684(14) & 0.317(14) & 0.322(8) \\
     & 0.010824 & 0.0155(3) & 0.0225(9) & 0.0510(10) & 0.684(16) & 0.312(19) & 0.321(9) \\
     & 0.007216 & 0.0107(5) & 0.0184(11) & 0.0495(14) & 0.683(19) & 0.305(26) & 0.322(11)
\end{tabular}
\end{table}

% figure 1

% figure 2

% figure 3

% figure 4

% figure 5

% figure 6

% figure 7

% figure 8

% figure 9

% figure 10

% figure 11

% figure 12

\end{document}